\documentclass[prd,aps,showpacs,nofootinbib,preprint,eqsecnum]{revtex4}
\usepackage{graphicx,color,amsmath,amsxtra}
\usepackage{epsf}
\usepackage{amssymb}
\usepackage{enumerate}
\usepackage{hhline}
\usepackage{array}
\usepackage{tabularx}
\usepackage{hangcaption}

\newcommand{\newc}{\newcommand}
\newc{\be}{\begin{equation}}
\newc{\ee}{\end{equation}}
\newc{\beq}{\begin{eqnarray}}
\newc{\eeq}{\end{eqnarray}}

\newcommand{\Eqn}[1]{&\hspace{-0.2em}#1\hspace{-0.2em}&}
\def\Vec#1{\mbox{\boldmath $#1$}}
\def\Vecs#1{\mbox{\boldmath\tiny $#1$}}

\begin{document}

\title{Large-scale magnetic fields from inflation due to a $CPT$-even 
Chern-Simons-like term with Kalb-Ramond and scalar fields
}
\author{
Kazuharu Bamba$^{1,2,3,}$\footnote{E-mail address: bamba@kmi.nagoya-u.ac.jp}, 
C. Q. Geng$^{1,4,}$\footnote{E-mail address: geng@phys.nthu.edu.tw}, 
S. H. Ho$^{5,6,}$\footnote{E-mail address: shho@mail.nctu.edu.tw, shho@mit.edu} 
and 
W. F. Kao$^{5,}$\footnote{E-mail address: gore@mail.nctu.edu.tw}}
\affiliation{
$^1$Department of Physics, National Tsing Hua University, Hsinchu, Taiwan 300\\
$^2$Kobayashi-Maskawa Institute for the Origin of Particles and the 
Universe, Nagoya University, Nagoya 464-8602, Japan\\
$^3$Eurasian International Center
for Theoretical Physics, Eurasian National University, Astana
010008, Kazakhstan\\
$^4$Physics Division, National Center for Theoretical Sciences, Hsinchu, Taiwan 300\\
$^5$Institute of Physics, National Chiao Tung University, Hsinchu, Taiwan 300\\
$^6$Center for Theoretical Physics, Massachusetts Institute of Technology,
Cambridge, MA 02139, USA} 

%\date{\today}

%%%%%%%%%%%%%%%%%%%%%
%  Abstract
%%%%%%%%%%%%%%%%%%%%%
\begin{abstract}

We investigate the generation of large-scale magnetic fields 
due to the breaking of the conformal invariance in the electromagnetic 
field through the $CPT$-even dimension-six Chern-Simons-like effective 
interaction with a fermion current by taking account of the 
dynamical Kalb-Ramond and scalar fields 
in inflationary cosmology. 
It is explicitly demonstrated that the magnetic fields on 1Mpc scale with the 
field strength of $\sim 10^{-9}$G at the present time can be induced. 

\end{abstract}
%%%%%%%%%%%%%%%%%%%%%

%----------------------------
%\pacs{
%Keywords:
%}
%\preprint{}
%\hspace{13.0cm}
%----------------------------

\maketitle
%==============================================================================

%%%%%%%%%%%%%%%%%%%%%%%%%%%
%%%  Sec. I
%%%%%%%%%%%%%%%%%%%%%%%%%%%
\section{Introduction}

Magnetic fields with the field strength 
$10^{-7}$--$10^{-6}$G on 10kpc--1Mpc scale 
in clusters of galaxies as well as 
$\sim 10^{-6}$G on 1--10kpc scale in galaxies of 
all types and at cosmological distances 
are observed 
(for reviews on the cosmic magnetic fields, 
see~\cite{reviews, Kandus:2010nw}). 
However, 
the origin of the cosmic magnetic fields, in particular 
the large-scale magnetic fields in clusters of galaxies, 
is not well understood yet. 
It is known that 
the dynamo amplification mechanism~\cite{EParker} can 
amplify very weak seed magnetic fields up to $\sim 10^{-6}$G 
in spiral galaxies, but 
its effectiveness in galaxies at high redshifts and clusters of galaxies 
is not confirmed. 
Other mechanisms to generate the cosmic magnetic fields exist, 
such as 
astrophysical processes~\cite{Biermann1, PI}, 
cosmological phase transitions~\cite{PT} and primordial density perturbations 
before or at the epoch of recombination~\cite{DP}. 
However, 
it is difficult for these mechanisms 
to induce the magnetic fields on megaparsec scales with sufficient field 
strengths to explain these observed in galaxies and 
clusters of galaxies without the dynamo amplification mechanism. 

It is considered that 
the most natural origin of the large-scale magnetic fields is from 
electromagnetic quantum fluctuations 
at the inflationary stage~\cite{Turner:1987bw}\footnote{ 
The back reaction of the generated magnetic fields on inflation 
has been argued~\cite{Backreaction}.
}. 
The reason is that 
inflation naturally extends the scale of the electromagnetic quantum 
fluctuations to the one larger than the Hubble horizon. 
The conformal invariance in the electromagnetic field must have been broken 
at the inflationary stage in order 
for electromagnetic quantum fluctuations to be produced 
during inflation~\cite{Parker:1968mv}\footnote{ 
The effect of the breaking of the conformal flatness 
due to scalar metric perturbations at 
the end of inflation has also been discussed~\cite{Maroto:2000zu}.
}. 
This is because the ordinary Maxwell theory is 
conformally invariant, whereas the metric is conformally flat 
in the Friedmann-Lema\^{i}tre-Robertson-Walker (FLRW) space-time. 
It should be cautioned that 
this does not apply if the FLRW background has nonzero spatial 
curvature such as an open FLRW background~\cite{NZC}. 
Various mechanisms of the breaking conformal invariance in the 
electromagnetic field have been proposed in the literature, 
such as those due to the non-minimal gravitational 
coupling~\cite{Turner:1987bw, NGC}, 
%%%
Weyl-Maxwell fields coupling~\cite{Tsagas:2009cr}, 
%%%
coupling to a scalar field~\cite{Ratra:1991bn, DE, C-S, pseudoscalar, Axions-Giovannini-B-BGH, charged scalar, ScalarED, ScalarED-2, Enqvist:2004yy, DBI}, 
%%%
generic coupling to a time-dependent background field~\cite{BS-B}, 
nonlinear electrodynamics~\cite{nonlinear-electrodynamics}, 
photon-graviphoton mixing~\cite{Gasperini:2000tw}, 
gravitoelectromagnetic inflationary formalism~\cite{Gravitoelectromagnetic}, 
%%%
conformal anomaly induced by quantum effects~\cite{Dolgov:1993vg}, 
spontaneous breaking of the Lorentz invariance~\cite{Bertolami:1998dn}, 
%%%
Lorentz violating term~\cite{L-V-T}, 
%%%
Lorentz gauge-breaking term~\cite{Jimenez:2010hu}, 
%%%
noncommutative field theory~\cite{NC}, 
preferred minimal length~\cite{Ashoorioon:2004rs}, 
cosmic defect~\cite{Hollenstein:2007kg}, 
bouncing cosmology~\cite{Salim:2006nw}, 
and Ho\v{r}ava-Lifshitz gravity~\cite{Maeda:2009hy}. 
For other breaking mechanisms and references, see a recent 
review in Ref.~\cite{Kandus:2010nw}. 
%%%%%%%%
Moreover, 
the complementary studies of magnetic catalysis in the gauge Higgs-Yukawa model and neutrino propagation in a strongly magnetized medium, also in view to their cosmological impact, have also been studied in Refs.~\cite{Elizalde:2002ca, Elizalde:2004mw, Elizalde:2000vz}. 
In addition, it is interesting to mention that 
a lower bound on the magnetic field strength in the hot universe has been recently obtained in Ref.~\cite{Elizalde:2012kz}. 
%%%%%%%%

% 
Recently, the $CPT$-even dimension-six Chern-Simons-like 
effective interaction between a fermion current and the electromagnetic field 
in inflationary cosmology 
has been studied to induce the cosmological 
birefringence~\cite{Geng:2007va, GHN2}, 
baryon number asymmetry~\cite{BGH}, 
and large-scale magnetic field~\cite{Bamba:2008hr}, respectively. 
%%%%%
In a related work~\cite{Campanelli:2008tt}, 
the generation of 
large-scale magnetic fields during inflation was examined 
in a Lorentz violating theory of Electrodynamics 
due to a Chern-Simons term coupling the 
$U(1)$ gauge field to an external four-vector, 
proposed in Ref.~\cite{Carroll:1989vb}. 
%%%%%
%%%%%
Furthermore, 
the $CPT$-even dimension-six Chern-Simons-like term 
with including the dynamical Kalb-Ramond and scalar fields was investigated 
to produce the cosmological birefringence~\cite{Ho:2010aq}. 
%%%%%
Spectral dependence of the cosmic microwave background (CMB) polarization 
and parity has also been discussed in Ref.~\cite{Balaji:2003sw}. 
%%%%%
The estimation of relic magnetic fields from CMB temperature correlations has 
been executed in Ref.~\cite{CMB-anisotropies-Giovannini}. 
Moreover, 
cosmological consequences of the existence of a Kalb-Ramond field 
have been studied in Ref.~\cite{Kalb-Ramond-field}. 
In addition, 
%searches for 
the role of spin and polarization in gravity 
have been considered in Ref.~\cite{Ni:2009fg} 
and 
limits on cosmological birefringence from the UV polarization of distant 
radio galaxies have been examined 
in Ref.~\cite{Limits-on-Cosmological-Birefringence}. 
%%%%% 
To search other cosmological ingredients from this term, 
in this paper we explore 
the generation of large-scale magnetic fields 
due to the breaking of the conformal invariance in the electromagnetic 
field through the $CPT$-even dimension-six Chern-Simons-like 
effective interaction with a fermion current 
by taking account of the dynamical Kalb-Ramond and scalar fields 
in inflationary cosmology. 

The paper is organized as follows.  
In Sec.\ II, we describe our model and derive equations of motion for 
the $U(1)$ gauge field. 
In Sec.\ III, we consider the evolution of the $U(1)$ gauge field 
and estimate the present strength of the large-scale magnetic fields. 
Finally, conclusions are given in Sec.\ IV.

%%%%%%%%%%%%%%%%%%%%%%%%%%%
%%%  Sec. II 
%%%%%%%%%%%%%%%%%%%%%%%%%%%
\section{The model}

Our model action is given by~\cite{Ho:2010aq}
\beq \label{action}
S
%S_0
%&=&
\Eqn{=} 
\int d^4x \sqrt{g} \biggl[-\frac{1}{2} \epsilon \phi^2 R - \frac{1}{2}  g^{\mu\nu} \partial_{\mu} \phi \partial_{\nu} \phi -V(\phi) \nonumber  \\
& & 
{}- \frac{\xi_1}{6\phi^2} \tilde{H}_{\mu\nu\alpha} \tilde{H}^{\mu\nu\alpha} +\frac{\xi_2}{\phi^2} j_{\mu}\left( A_{\nu}\tilde{F}^{\mu\nu}+ \frac{1}{2}  \epsilon^{\mu\nu\alpha\beta} \partial_{\nu} B_{\alpha\beta}\right) -\frac{1}{4}  
F^{\mu\nu}F_{\mu\nu}\biggr]\,, 
\eeq
where $g$ is the determinant of the metric tensor $g_{\mu\nu}$, 
$F_{\mu\nu} = \partial_{\mu}A_{\nu}-\partial_{\nu}A_{\mu}$ is 
the electromagnetic field strength tensor, 
$\tilde{F}^{\mu\nu}=(1/2)\epsilon^{\mu\nu\alpha\beta}F_{\alpha\beta}$
is the dual of $F_{\mu\nu}$
with $\epsilon^{\mu\nu\alpha\beta}=(1/\sqrt{g})e^{\mu\nu\alpha\beta}$ being 
the Levi-Civita tensor normalized by $e^{0123}=+1$, $R$ is the Ricci scalar, 
$\phi$ is a dynamical scalar field, and $B_{\mu\nu}$ is the Kalb-Ramond fields 
with the modified field strength $\tilde{H}_{\mu\nu\alpha} = 
\partial_{[\mu}B_{\nu\alpha]}+A_{[\mu}F_{\nu\alpha]}$. 
%%%%%
We use units 
of $k_\mathrm{B} = c = \hbar = 1$ and adopt 
Heaviside-Lorentz units of electromagnetism. 
%%%%%

The following set of equations of motion can be obtained 
by varying the action with respect to $\phi$, $g_{\mu\nu}$, $B_{\mu\nu}$ 
and $A_{\mu}$: 
\beq \label{vary phi}
\epsilon \phi R = D_{\mu} \partial^{\mu}\phi - \frac{\partial V}{\partial \phi} + \frac{\xi_1}{3\phi^3} \tilde{H}^2 -2 \frac{\xi_2}{\phi^3} j_{\mu}\left( A_{\nu}\tilde{F}^{\mu\nu}+ \frac{1}{2}  \epsilon^{\mu\nu\alpha\beta} \partial_{\nu} B_{\alpha\beta}\right)\,, 
\eeq
\beq \label{vary g}
\epsilon \phi^2 G_{\mu\nu} 
%&= &
\Eqn{=}
\left[\frac{1}{2} \left(\partial_{\alpha}\phi\right)^2+V(\phi)\right]g_{\mu\nu} -\partial_{\mu}\phi \partial_{\nu} \phi + \frac{\xi_1}{6\phi^2} \tilde{H}^2 g_{\mu\nu} +\left(\frac{1}{4} F^2 g_{\mu\nu} - F_{\mu\alpha}F_{\nu}^{\alpha} 
\right) 
\nonumber \\
%&-&
& &
{}+ \epsilon(D_{\nu}D_{\mu}\phi^2-D^{\sigma}D_{\sigma}\phi^2 g_{\mu\nu}) 
-\frac{1}{\phi^2}
\tilde{H}_{\mu\alpha\beta}\tilde{H}_{\nu}^{\alpha\beta}\,, 
\eeq
\beq \label{vary B}
D_{\mu}\left(\frac{\xi_1}{\phi^2}\tilde{H}^{\mu\nu\alpha}+\frac{\xi_2}{2\phi^2}\epsilon^{\mu\nu\alpha\beta}j_{\beta}\right)=0\,,
\eeq
\beq \label{vary A}
D_{\nu}F^{\nu\mu}-D_{\nu}\left(\frac{2\xi_1}{\phi^2}\tilde{H}^{\nu\alpha\mu}A_{\alpha}+\frac{\xi_2}{\phi^2}\epsilon^{\beta\alpha\nu\mu}j_{\beta}A_{\alpha}
\right)=\frac{\xi_1}{\phi^2}\tilde{H}^{\mu\nu\alpha}F_{\nu\alpha}-\frac{\xi_2}{\phi^2}j_{\nu}\tilde{F}^{\nu\mu}\,.
\eeq
Since $\tilde{H}^{\mu\nu\alpha}$ is a totally antisymmetric tensor, 
we can write 
$\tilde{H}^{\mu\nu\alpha}=\epsilon^{\mu\nu\alpha\beta} T_{\beta}$, 
where $T_{\beta}$ is a vector with mass dimension three. 
Thus, Eq.~(\ref{vary B}) is rewritten to 
\beq \label{torsion curl}
%
%D_{\mu}\left(\frac{\xi_1}{\phi^2}\epsilon^{\mu\nu\alpha\beta} T_{\beta}+\frac{\%xi_2}{2\phi^2}\epsilon^{\mu\nu\alpha\beta} j_{\beta}\right)=0 \nonumber \\
%\Rightarrow 
%
\epsilon^{\mu\nu\alpha\beta} \partial_{\mu}\left(\frac{\xi_1}{\phi^2}T_{\beta}+\frac{\xi_2}{2\phi^2} j_{\beta}\right)=0\,. 
\label{eq:A-10}
\eeq
Focusing on the space-time manifold with the first trivial homology group, any closed one-form is an exact one-form. Therefore, 
from Eq.~(\ref{eq:A-10}), we can 
express the torsion field as 
\beq \label{torsion}
%\phi^{-2}
\frac{1}{\phi^{2}}
\left(\xi_1 T_{\beta}+\frac{\xi_2}{2} j_{\beta}\right)=
\partial_{\beta} 
%\Lambda
\Phi\,, 
\eeq
where 
$\Phi$ is a dimensionless pseudo-scalar. With the help of 
Eq.~(\ref{torsion}), we can further simplify the equations of motion 
(\ref{vary phi}), (\ref{vary g}) and (\ref{vary A})
to be 
\beq 
%\label{eom phi}
\epsilon \phi R
%&=&
%\Eqn{=} 
%D_{\mu} \partial^{\mu}\phi - \frac{\partial V}{\partial \phi}+\frac{1}{3\xi_1\phi^3}(\frac{-6}{\xi_1^2})\left[\phi^4\left(\partial_{\mu}\Phi\right)^2-\xi_2 %\phi^2
%(\partial){\mu} 
%\left(\partial_{\mu} 
%\Phi \right)j^{\mu}+\frac{\xi_2^2}{4} j_{\mu} j^{\mu}\right] \nonumber \\
%&-&
%& &
%{}-
%\frac{2\xi_2}{\phi^3} j_{\mu}\left[A_{\nu}\tilde{F}^{\mu\nu}+\left(\phi^2 \partial^{\mu} \Phi-\frac{\xi_2}{2} j^{\mu}\right)-\frac{1}%{2}\epsilon^{\mu\nu\alpha\beta} A_{\beta} F_{\nu\alpha}\right] \nonumber \\
%&=& 
\Eqn{=} 
D_{\mu} \partial^{\mu} \phi -\frac{\partial V}{\partial \phi}-\frac{2\phi}{3\xi_1}\left(\partial_{\mu} \Phi\right)^2+\frac{\xi_2^2}{2\xi_1 \phi^3}\left(j_{\mu}\right)^2\,, 
\label{eom phi} \\
%\eeq
%
%\beq 
\label{eom g}
\epsilon \phi^2 G_{\mu\nu}
%&=&
\Eqn{=} 
\left[\frac{1}{2}\left(\partial_{\alpha} \phi\right)^2+V(\phi)\right]g_{\mu\nu}-\partial_{\mu}\phi \partial_{\nu}\phi +\epsilon(D_{\nu}D_{\mu}\phi^2-D^{\sigma}D_{\sigma}\phi^2 g_{\mu\nu}) \nonumber \\
% &+&
& &
{}+
\frac{1}{\xi_1 \phi^2} \left[\phi^4\left(\partial_{\alpha}\Phi\right)^2-\xi_2 
\phi^2 j_{\alpha} \partial^{\alpha} \Phi +\frac{\xi_2^2}{4}\left( j_{\mu} 
\right)^2 
\right] g_{\mu\nu} \nonumber \\
% &+& 
& &
{}+ 
\left(\frac{1}{4}F^2 g_{\mu\nu}-F_{\mu\alpha}F_{\nu}^{\alpha}\right)-2\frac{\xi_1}{\phi^2}
\left(\frac{\phi^2}{\xi_1}\partial_{\mu}\Phi - \frac{\xi_2}{2\xi_1}j_{\mu}\right) \left(\frac{\phi^2}{\xi_1}\partial_{\nu}\Phi - \frac{\xi_2}{2\xi_1}j_{\nu}\right) \,,  
\label{eom g} \\
%\eeq
%
%\beq 
%\label{eom A}
D_{\mu}F^{\mu\nu}
%=
\Eqn{=} 
-4 \left(\partial_{\mu} 
\Phi\right) \tilde{F}^{\mu\nu}\,, 
\label{eom A}
\eeq
respectively. 

Now, we consider the simplest $\phi^4$ potential, given by 
\beq \label{potential}
V(\phi)
%=-m^2\phi^2+\lambda \phi^4=\lambda\left(\phi^2-\frac{m^2}{2\lambda}\right)^2-\frac{m^4}{4\lambda}\,.
=\lambda\left(\phi^2-\phi_0^2\right)^2+V_0\,, 
\eeq
where $V_0$ and $\lambda$ are both larger than zero. 
%%%%%
We take the flat FLRW space-time with the metric, 
\begin{eqnarray}
{ds}^2 = -{dt}^2 + a^2(t)d{\Vec{x}}^2\,,
\label{eq:3.1}
\end{eqnarray}
where $a(t)$ is the scale factor. 
%%%%%
In the FLRW universe, it is reasonable to 
assume a homogeneous and isotropic fermion current and Kalb-Ramond 
field~\cite{Ho:2010aq}, i.e., 
$j_{\mu}=( j_0 (t), \Vec{0})$ and 
$T_{\mu}=( T_0 (t), \Vec{0})$. 
{}From Eqs.~(\ref{eom phi}) and (\ref{eom g}), 
we have 
\beq \label{sol lambda}
\partial_0 \Phi =-2\frac{\xi_1V_0}{\xi_2 j_0}+\frac{\xi_2 }{2\phi_0^2}j_0\,. 
\eeq
For the Coulomb gauge of 
$A_0(t,\Vec{x}) = 0$ and ${\partial}_j A^j (t,\Vec{x}) =0$, 
Eq.~(\ref{eom A}) becomes 
\beq \label{eq:2.14}
%\label{A}
&& \ddot{A_j}(t,\Vec{x})+H\dot{A_j}(t,\Vec{x})
-\frac{1}{a^2}\partial_i\partial_i A_j(t,\Vec{x})-4\frac{\dot{\Phi}}{a}e_{jik}\partial_i A_k(t,\Vec{x})=0\,,
\eeq
where a dot denotes a time derivative, $H=\dot{a}/a$ is 
the Hubble parameter, and $e_{ijk}$ is the totally antisymmetric 
tensor $(e_{123} = +1)$.

%%%%%%%%%%%%%%%%%%%%%%%%%%%
%%%  Sec. III 
%%%%%%%%%%%%%%%%%%%%%%%%%%%
\section{Large-scale magnetic fields}

%%%%%%%%%%%%%%%%%%%%%%%%%%%
%%%  Sec. III A
%%%%%%%%%%%%%%%%%%%%%%%%%%%
\subsection{Evolution of the $U(1)$ gauge field}

We consider the case in which a slow-roll exponential inflation 
occurs with 
$
a(t) = a_1 \exp \left[ H_{\mathrm{inf}}(t-t_1) \right], 
$
where $a_1$ is the scale factor at the time $t_1$ when a 
comoving wavelength $2\pi/k$ of the $U(1)$ gauge field 
first crosses outside the horizon during 
inflation, $k/(a_1 H_{\mathrm{inf}}) = 1$, and 
$H_{\mathrm{inf}}$ is the Hubble constant at the inflationary stage. 

%%%%%
It follows from the quantization of 
the $U(1)$ gauge field $A_{\mu}(t,\Vec{x})$ 
that $A_i(t,\Vec{x})$ is expressed as 
\begin{equation}
A_i(t,\Vec{x}) = 
\int \frac{d^3 k}{{(2\pi)}^{3/2}} 
\biggl[ \hat{b}(\Vec{k}) A_i(t,\Vec{k})e^{i \Vecs{k} \cdot \Vecs{x} } 
+{\hat{b}}^{\dagger}(\Vec{k}) 
{A_i^*}(t,\Vec{k})e^{-i \Vecs{k} \cdot \Vecs{x}} \biggr]\,,
\label{eq:8} 
\end{equation}
where $\Vec{k}$ is the comoving wave number, $k$ denotes its amplitude 
$|\Vec{k}|$, and $\hat{b}(\Vec{k})$ and ${\hat{b}}^{\dagger}(\Vec{k})$ 
are the annihilation and creation operators which satisfy 
$
\left[ \hat{b}(\Vec{k}), {\hat{b}}^{\dagger}({\Vec{k}}^{\prime}) \right] = 
{\delta}^3 (\Vec{k}-{\Vec{k}}^{\prime})
 \hspace{1mm}
\mathrm{and}\ \mathrm{others} = 0
$. 
%%%%%
In what follows, we choose the $x^3$ axis to lie along the spatial momentum 
direction \Vec{k} and denote the transverse directions $x^{I}$ with 
$I=1, 2$. 
We use circular polarizations expressed by 
the combination of linear polarizations as 
$A_{\pm}(k,t) \equiv A_1(k,t) \pm i A_2(k,t)$. 
{}From Eq.\ (\ref{eq:2.14}), we obtain 
%
%\beq 
\be
\label{eom A-1}
\ddot{A}_{\pm}(k,t)+H\dot{A}_{\pm}(k,t)+\frac{k^2}{a^2}A_{\pm}(k,t) 
\mp 4\dot{\Phi}\frac{k}{a}A_{\pm}(k,t)=0\,,
%\eeq
\ee
with 
%
%\beq 
\be
\label{Phi}
\dot{\Phi}=-\frac{\xi_1}{\xi_2} \frac{2V_0}{j_0}+\frac{\xi_2}{2\phi_0^2}j_0\,, 
%\eeq
\ee
where $j_0$ is the fermion number density. 
Here, we take 
$j_0=\bar{n}a^{-m}$ ($m>0$), $\phi^2=M_{\mathrm{Pl}}^2=1/(8\pi G)$, 
$H_{\mathrm{inf}}=10^{10}$GeV, and 
the comoving scale $L=2\pi /k=1$Mpc. 
Since there exists no analytic solution of Eq. (\ref{eom A-1}), 
we investigate the numerical solutions for 
$m=1$, $2$ and $3$. 
Note that $m=3$ corresponds to the conventional property. 

By using $k/a=(k/a_1) e^{-H_{\mathrm{inf}}(t-t_1)}=H_{\mathrm{inf}}e^{-H_{\mathrm{inf}}(t-t_1)}$, 
the equation of motion (\ref{eom A-1}) is rewritten to 
\beq \label{eom B}
&& \hspace{-5mm}
\ddot{A}_{\pm}(k,t)+H_{\mathrm{inf}}\dot{A}_{\pm}(k,t) \nonumber \\
&& \hspace{-5mm}
+\left[H_{\mathrm{inf}}^2 e^{-2H_{\mathrm{inf}}(t-t_1)}\mp 4 H_{\mathrm{inf}}e^{-H_{\mathrm{inf}}(t-t_1)}\left(-\frac{\xi_1}{\xi_2}\frac{2V_0}{\bar{n}}a^m+\frac{\xi_2}{2M_{\mathrm{Pl}}^2}\bar{n}a^{-m}\right)\right]A_{\pm}(k,t)=0\,, 
\eeq
which can be simplified to be 
\beq \label{eom C}
A_{\pm}^{\prime\prime}(k,\tilde{t}\,)+A_{\pm}^{\prime}(k,\tilde{t}\,)
+e^{-2(\tilde{t}-\tilde{t}_1)} 
\left[1\mp J_1 e^{(m+1)(\tilde{t}-\tilde{t}_1)} 
\mp J_2 e^{(1-m)(\tilde{t}-\tilde{t}_1)} \right] A_{\pm}(k,\tilde{t}\,)=0\,, 
\eeq
where the prime denotes a derivative with respect to $\tilde{t}$ and 
\beq
&&\tilde{t} \equiv H_{\mathrm{inf}}t\,, \\
&&J_1 \equiv \frac{4}{H_{\mathrm{inf}}}\left[-\frac{\xi_1}{\xi_2}\frac{2V_0}{\bar{n}}\left(\frac{k}{H_{\mathrm{inf}}}\right)^m\right]\,, \\
&&J_2 \equiv \frac{4}{H_{\mathrm{inf}}}\left[\frac{\xi_2}{2M_{\mathrm{Pl}}^2}\bar{n}\left(\frac{k}{H_{\mathrm{inf}}}\right)^{-m}\right]\,. 
\eeq

We take the following initial conditions at 
$\tilde{t}_1= H_{\mathrm{inf}}t_1 =1$: 
\beq \label{initial}
A_{\pm}(k,\tilde{t}_1=1)=\frac{1}{\sqrt{2k}}\,, \quad 
A^{\prime}_{\pm}(k,\tilde{t}_1=1)=\frac{H_{\mathrm{inf}}}{\sqrt{2k}}\,, 
\eeq
in order that the vacuum should reduce to the one in Minkowski space-time 
in the short-wavelength limit. 
For convenience in numerical calculations, we introduce the variable 
$C_{\pm}(k,\tilde{t}\,)$ to separate 
the coefficient $1/\sqrt{2k}$ from the amplitudes $A_{\pm}(k,\tilde{t}\,)$ as 
%
%\beq \label{eom D}
%&& 
$
A_{\pm}(k,\tilde{t}\,) 
%\Eqn{=} 
=
C_{\pm}(k,\tilde{t}\,) A_{\pm}(k, \tilde{t}_1)=
%\frac{1}{\sqrt{2k}}
\left(1/\sqrt{2k}\right) 
C_{\pm} (k,\tilde{t}\,)
$ 
and 
%\nonumber 
%\\
%&&
$
A^{\prime}_{\pm}(k,\tilde{t}\,) 
%\Eqn{=} 
= 
C_{\pm}^{\prime}(k,\tilde{t}\,) A_{\pm}(k, \tilde{t}_1)=
%\frac{1}{\sqrt{2k}} 
\left(1/\sqrt{2k}\right) 
C_{\pm}^{\prime}(k,\tilde{t}\,) 
$. 
%\nonumber 
%\eeq
%
{}From Eq.~(\ref{eom C}), we find 
\beq \label{eq310}
C_{\pm}^{\prime\prime}(k,\tilde{t}\,)
+C_{\pm}^{\prime}(k,\tilde{t}\,)+e^{-2(\tilde{t}-\tilde{t}_1)} 
\left[1\mp J_1 e^{(m+1)(\tilde{t}-\tilde{t}_1)} \mp 
J_2 e^{(1-m)(\tilde{t}-\tilde{t}_1)} \right] C_{\pm}(k,\tilde{t}\,)=0\,,
\eeq
with the initial conditions at $\tilde{t} = 1$ as 
%
%\beq
$C_{\pm}(k,\tilde{t}=1)=1$ 
and 
$C_{\pm}^{\prime}(k,\tilde{t}=1) =H_{\mathrm{inf}}$. 
%\eeq
%

\begin{center}
\begin{figure}[tbp]
\begin{tabular}{ll}
\begin{minipage}{80mm}
\begin{center}
\unitlength=1mm
\resizebox{!}{6.5cm}{
   \includegraphics{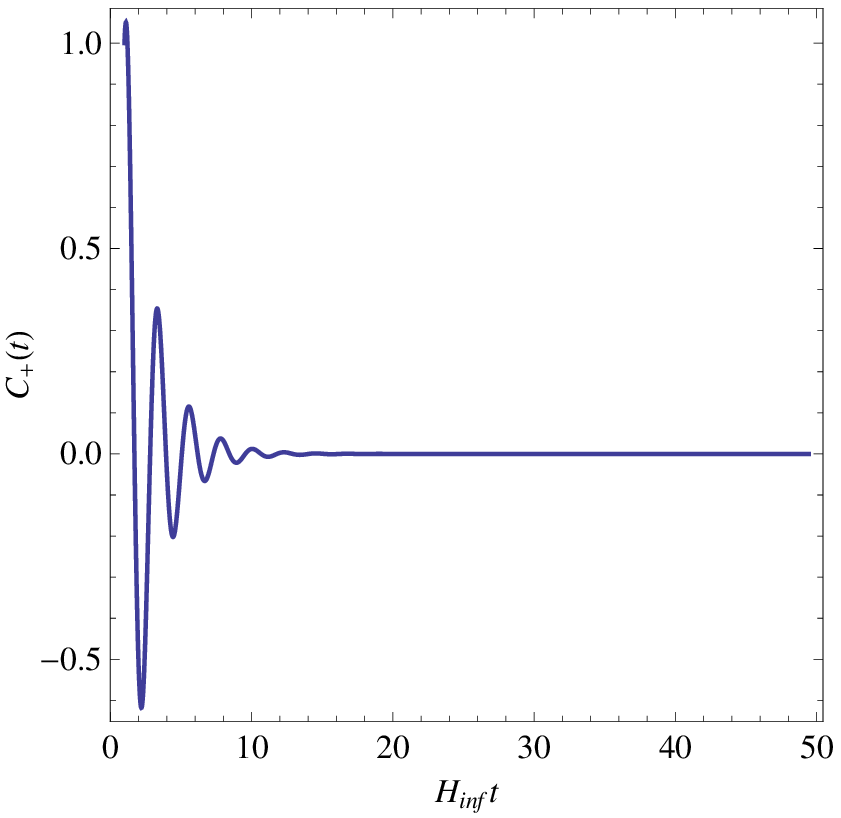}
                  }
\end{center}
\end{minipage}
&
\begin{minipage}{80mm}
\begin{center}
\unitlength=1mm
\resizebox{!}{6.5cm}{
   \includegraphics{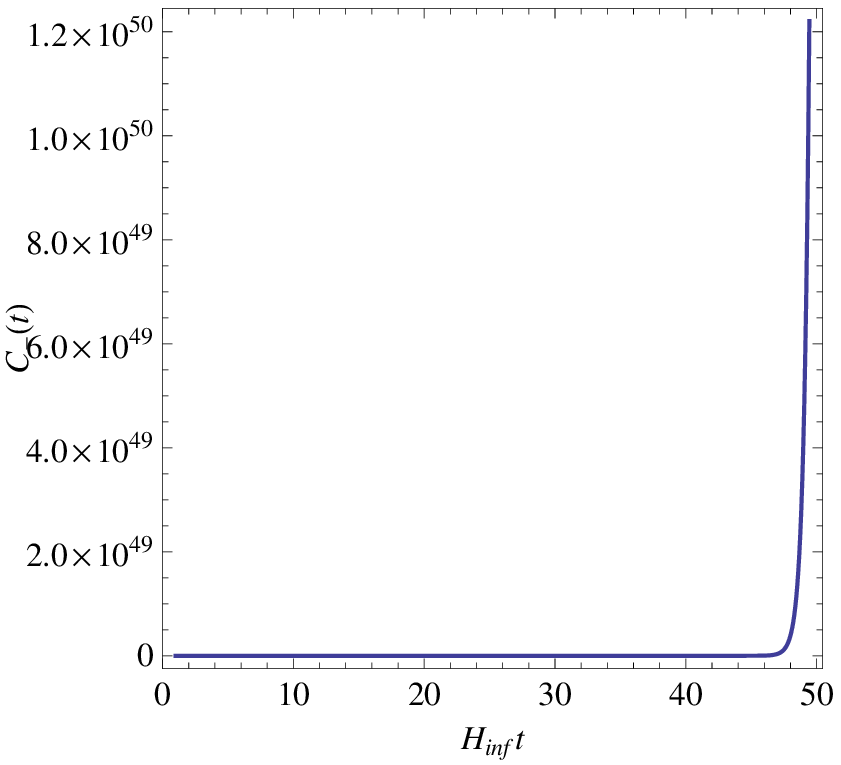}
                  }
\end{center}
\end{minipage}

\end{tabular}
\caption{$C_+(k,\tilde{t}\,)$ (left) and $C_-(k,\tilde{t}\,)$ 
(right) as functions of $\tilde{t} \equiv H_{\mathrm{inf}}t$ 
with a comoving scale $L=2\pi/k = 1\mathrm{Mpc}$ 
for $n=\bar{n}a^{-1}$, where $\bar{n}=10^{-104.36}$, 
$H_{\mathrm{inf}} = 10^{10}$GeV, $V_0=10^{-47} \ \mathrm{GeV}^4$ 
and $\xi_1=\xi_2=1$. 
}
\label{fig-1}
\end{figure}
\end{center}

\begin{center}
\begin{figure}[tbp]
\begin{tabular}{ll}
\begin{minipage}{80mm}
\begin{center}
\unitlength=1mm
\resizebox{!}{6.5cm}{
   \includegraphics{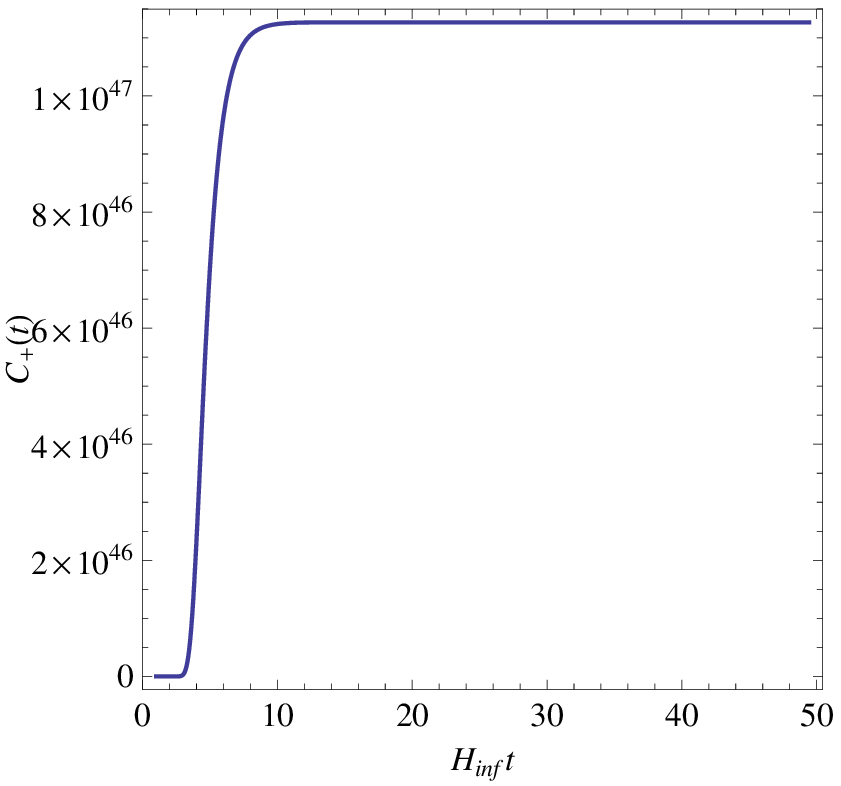}
                  }
\end{center}
\end{minipage}
&
\begin{minipage}{80mm}
\begin{center}
\unitlength=1mm
\resizebox{!}{6.5cm}{
   \includegraphics{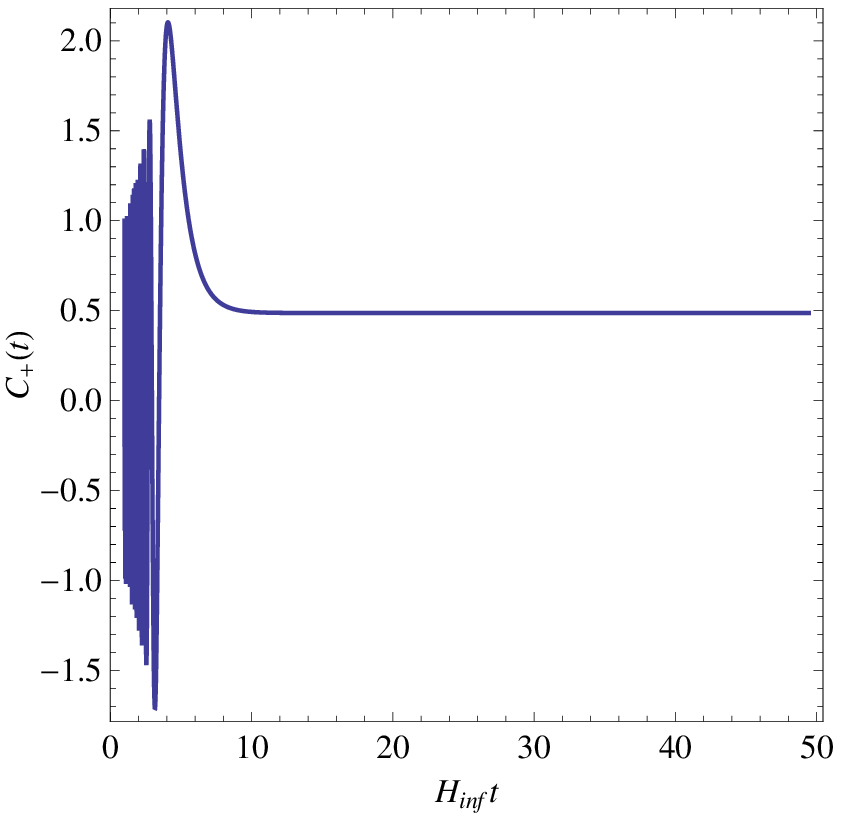}
                  }
\end{center}
\end{minipage}

\end{tabular}
\caption{Legend is the same as Fig.~1 but for $n=\bar{n}a^{-2}$ with 
$\bar{n}=10^{-45.3}$. 
}
\label{fig-2}
\end{figure}
\end{center}

\begin{center}
\begin{figure}[tbp]
\begin{tabular}{ll}
\begin{minipage}{80mm}
\begin{center}
\unitlength=1mm
\resizebox{!}{6.5cm}{
   \includegraphics{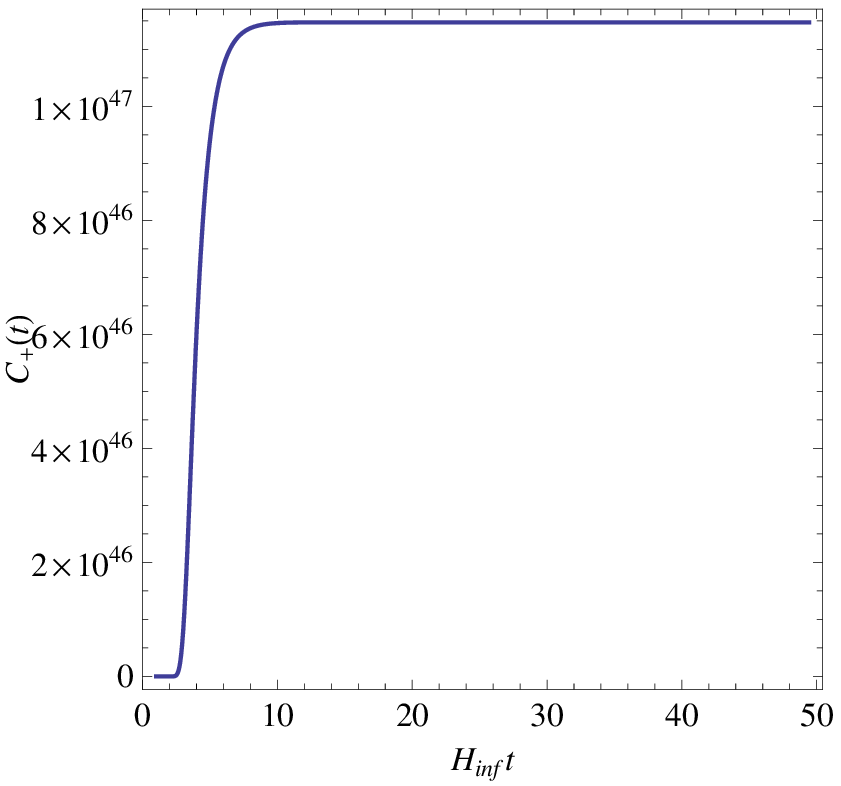}
                  }
\end{center}
\end{minipage}
&
\begin{minipage}{80mm}
\begin{center}
\unitlength=1mm
\resizebox{!}{6.5cm}{
   \includegraphics{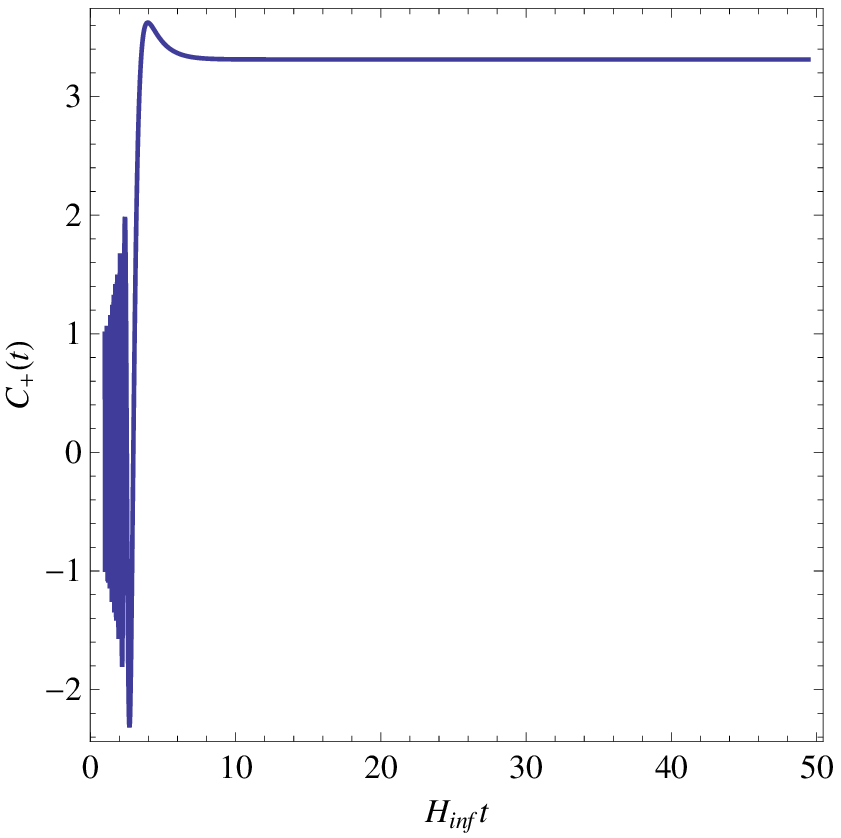}
                  }
\end{center}
\end{minipage}

\end{tabular}
\caption{Legend is the same as Fig.~1 but for $n=\bar{n}a^{-3}$ with 
$\bar{n}=10^{-92.45}$. 
}
\label{fig-3}
\end{figure}
\end{center}

%%%%%
In Figs.~\ref{fig-1}, \ref{fig-2} and \ref{fig-3}, we depict 
$C_+(k,\tilde{t}\,)$ (left) and $C_-(k,\tilde{t}\,)$ 
(right) as functions of $\tilde{t} \equiv H_{\mathrm{inf}}t$ 
with a comoving scale $L=2\pi/k = 1\mathrm{Mpc}$ 
for $n=\bar{n}a^{-1}$ ($\bar{n}=10^{-104.36}$), 
$n=\bar{n}a^{-2}$ ($\bar{n}=10^{-45.3}$), 
and $n=\bar{n}a^{-3}$ ($\bar{n}=10^{-92.45}$), 
respectively, 
where 
$H_{\mathrm{inf}} = 10^{10}$GeV, $V_0=10^{-47} \ \mathrm{GeV}^4$ 
and $\xi_1=\xi_2=1$. 
%%%%%
It is known that $H_{\mathrm{inf}} < 6.0 \times 10^{14}$GeV 
from tensor perturbations~\cite{C-H} with the observational data 
on the anisotropy of the CMB 
%the cosmic microwave background (CMB) 
radiation~\cite{Komatsu:2010fb}. 

%%%%%
We note 
that the evolutions of $C_+(k,\tilde{t}\,)$ and $C_-(k,\tilde{t}\,)$ 
depend on $\bar{n}$. 
To generate the magnetic fields with enough strength,  
we have to choose a specific value of $\bar{n}$. 
We also remark that the behaviors of $C_+(k,\tilde{t}\,)$ and 
$C_-(k,\tilde{t}\,)$ for $m=1$ 
are different from those for $m>1$. 
In $m=1$, 
$C_+(k,\tilde{t}\,)$ 
approaches a constant at a large $\tilde{t}$, 
but $C_-(k,\tilde{t}\,)$ increases 
with $\tilde{t}$. 
On the other hand, 
for $m=2$ and $3$, 
both $C_+(k,\tilde{t}\,)$ and $C_-(k,\tilde{t}\,)$ 
become constants at large $\tilde{t}$. We will discuss the asymptotic behavior of $C_+(k,\tilde{t}\,)$ and $C_-(k,\tilde{t}\,)$ in Appendix \ref{A}.
%\ref{A1}.
%%%%%

%%%%%%%%%%%%%%%%%%%%%
%%%  Figures
%%%%%%%%%%%%%%%%%%%%%
%%%%%% Fig. 1 %%%%%%%%%
%%%%%%%%%%%%%%%%%%%%%%%%

%%%%%% Fig. 2 %%%%%%%%%
%%%%%%%%%%%%%%%%%%%%%%%%

%%%%%% Fig. 3 %%%%%%%%%
%%%%%%%%%%%%%%%%%%%%%%%%

%%%%%%%%%%%%%%
\vspace{-10mm}
%%%%%%%%%%%%%%

%%%%%%%%%%%%%%%%%%%%%%%%%%%
%%%  Sec. III B
%%%%%%%%%%%%%%%%%%%%%%%%%%%
\subsection{Strength of the large-scale magnetic fields}

We estimate the present strength of the large-scale magnetic fields 
by using the numerical results for $C_{\pm}(k,\tilde{t}\,)$. 
The proper magnetic fields are given by~\cite{Ratra:1991bn} 
\beq \label{proper}
B^{\mathrm{proper}}_i (t,x) = a^{-1}B_i (t,x) = a^{-2}\epsilon_{ijk} \partial_j A_k(t,x)\,,
\eeq
where ${B}_i(t,\Vec{x})$ are the comoving magnetic fields. 
The energy density in Fourier space is given by 
\beq \label{density}
&&\rho_B(k,t)=\frac{1}{2}\left[ |B^{\mathrm{proper}}_+(k,t)|^2 +|B^{\mathrm{proper}}_-(k,t)|^2\right]\,, 
%\nonumber 
\\
&& |B^{\mathrm{proper}}_{\pm}(k,t)|^2=\frac{1}{a^2}\left(\frac{k}{a}\right)^2 |A_{\pm}(k,t)|^2\,,
\eeq
where $B_\pm^{\mathrm{proper}}(k,t) \equiv 
B_1^{\mathrm{proper}}(k,t) \pm i B_2^{\mathrm{proper}}(k,t)$. 
Multiplying $\rho_B(k,t)$ by the phase space density of $4\pi k^3/(2\pi )^3$, 
we obtain the energy density of the proper magnetic field as 
%
%\beq 
\be 
\label{proper density}
%&&
\rho_B(L,t)=\frac{1}{8\pi^2}\left(\frac{k}{a}\right)^4 
%\emph{I}
\mathcal{I}(k,t)\,, 
%\nonumber 
%\\
\ee
with 
\be 
%&& 
%\emph{I} 
\mathcal{I}(k,t)=|C_+(k,t)|^2 +|C_-(k,t)|^2\,, 
%\eeq
\ee
where $\mathcal{I}(k,t)$ 
can be interpreted as the amplification factor 
at the inflationary stage. 
%
%%%%%
Here, 
we concentrate on the situation in which 
after inflation the universe is reheated immediately at 
$t=t_\mathrm{R}$. The conductivity of the universe ${\sigma}_\mathrm{c}$ 
is negligibly small during inflation because there are few charged particles 
at that time. 
After reheating, charged particles are produced 
so that the conductivity immediately jumps to a large value:\ 
${\sigma}_\mathrm{c} \gg H$. 
For a large enough ${\sigma}_\mathrm{c}$, 
magnetic fields evolve in proportion to $a^{-2}(t)$~\cite{Ratra:1991bn}. 
%%%%%
{}From $B(L,t_0)=\sqrt{2\rho_B(k,t_0)}$ and 
Eq.~(\ref{proper density}), we find that 
$B(L,t_0)=\left[1/\left(2\pi\right)\right] \left( k/a_\mathrm{R} \right)^2 
\left( a_\mathrm{R}/a_0\right)^2 
\sqrt{\mathcal{I}(k,t_\mathrm{R})}
$. 
As a result, 
the present strength of the magnetic fields is described as 
%
%\beq 
\be
\label{mag strength}
B(L,t_0)
%=&&\sqrt{2\rho_B(k,t_0)} \nonumber \\
%=&&\frac{1}{2\pi}\left( \frac{k}{a_0}\right)^2 
%\sqrt{
%%\emph{I}
%\mathcal{I}(k,t_\mathrm{R})} \nonumber \\
%=&&\frac{1}{2\pi}\left( \frac{k}{a_\mathrm{R}}\right)^2 
%\left( \frac{a_\mathrm{R}}{a_0}\right)^2\sqrt{
%%\emph{I}
%\mathcal{I}(k,t_\mathrm{R})}  \ \ [\mathrm{GeV}]\nonumber \\
=
%&&
\left(\frac{10^{20}}{1.95}\right)\frac{1}{2\pi}\left(\frac{k}{a_1}\right)^2 e^{-2N}\left(\frac{a_\mathrm{R}}{a_0}\right)^2
\sqrt{
%\emph{I}
\mathcal{I}(k,t_\mathrm{R})} \,\, [\mathrm{G}]\,,
%\ \ [\mathrm{Gauss}]\,,
%\eeq
\ee
where $ a_\mathrm{R}/a_0 = (g_R/3.91)^{-1/3}T_{\gamma 0}/T_\mathrm{R} $ with 
$T_\mathrm{R}$ being the reheating temperature and 
$
T_{ \gamma 0} \left( = 2.73 [\mathrm{K}] \right)
$ 
the present temperature of the CMB radiation~\cite{Kolb and Turner}, 
$a_\mathrm{R}$ and $a_0 (= 1)$ are the values of $a$ 
at $t = t_{\mathrm{R}}$ and the present time $t_0$, 
and $N$ is the number of $e$-folds 
between the time $t_1$ and $t_\mathrm{R}$, given by 
$N = 45 + \ln (L/[\mathrm{Mpc}]) + \ln \Xi$, 
where $\Xi=[30/(\pi^2 g_\mathrm{R})]^{1/12} 
\rho_\mathrm{R}^{1/4}/(10^{38/3}[\mathrm{GeV}])$, 
$g_\mathrm{R} \sim 100$ is the total number degree of freedom for relativistic 
particles at the reheating epoch, and 
$\rho_\mathrm{R}=(\pi^2/30)g_\mathrm{R} T_\mathrm{R}^4$ is the energy density 
of radiation at the reheating stage. 

Using Eq.\ (\ref{mag strength}) and 
$H_{\mathrm{inf}}^2 = 
\left(8\pi/3\right)\rho_{\mathrm{R}}/M_{\mathrm{Pl}}^2
$, 
we find that 
when 
$H_{\mathrm{inf}} = 10^{10}$GeV, $V_0=10^{-47} \ \mathrm{GeV}^4$ 
and $\xi_1=\xi_2=1$, 
the generated magnetic field on 
1Mpc scale at the present time is 
$B_0 (L=1\mathrm{Mpc}, t_0) = 
4.1 \times 10^{-9}$G, 
$1.7 \times 10^{-9}$G 
and 
$1.7 \times 10^{-9}$G 
for the cases in Figs.~\ref{fig-1}, \ref{fig-2} and \ref{fig-3}, 
respectively. 

%%%%%
%%%%%
Finally, we mention constraints on the primordial magnetic fields 
from the Big Bang Nucleosynthesis (BBN) 
and CMB anisotropy measurements on small and large\footnote{
There also exist constraints on the magnetic field strength on large scales 
from the matter density fluctuation parameter $\sigma_8$~\cite{YIKM}, 
the fifth science (S5) run of laser interferometer gravitational-wave 
observatory (LIGO)~\cite{Wang:2008vp}, 
and 
Chandra X-ray galaxy cluster survey as well as Sunyaev-Zel'divich (S-Z) 
survey~\cite{Tashiro:2010st}, 
which are consistent with or weaker than those from CMB. 
} scales, respectively\footnote{
Generic property of the spectrum of large-scale magnetic fields from inflation 
has been discussed~\cite{Bamba:2007hm}.}. 
The limit on the present strength of the magnetic fields around 
the BBN horizon 
size $\sim 9.8 \times 10^{-5} h^{-1}\mathrm{Mpc}$ 
with $h=0.7$~\cite{Freedman:2000cf} 
is less than 
$10^{-6}$G~\cite{BBN}. 
For the cases in Figs.~\ref{fig-2} and \ref{fig-3}, 
the present strength on the BBN horizon scale is 
$1.5 \times 10^{-48}$G 
and 
$1.5 \times 10^{-48}$G, 
respectively, 
which are consistent with the constraints from BBN, 
whereas for that in Fig.~\ref{fig-1}, 
it is diverging large. 
On the other hand, 
the result of $\sim 10^{-9}$G on 1Mpc scale 
for all cases in Figs.~\ref{fig-1}--\ref{fig-3} 
is consistent with the observational upper bounds 
($\sim 2-6 \times 10^{-9}$G) 
from CMB~\cite{CMB-Limit}\footnote{
The limit from CMB on the current strength 
on scales larger than the present horizon 
is less than $4.8 \times 10^{-9} \mathrm{G}$~\cite{Barrow:1997mj}. 
To satisfy this limit, for example, 
one may take 
$n=\bar{n}a^{-2}$ with $\bar{n}=10^{-52.44}$, 
$H_{\mathrm{inf}} = 10^{10}$GeV, $V_0=10^{-47} \ \mathrm{GeV}^4$ 
and $\xi_1=\xi_2=1$. 
In this case, 
the present strength of the magnetic fields 
is $1.4 \times 10^{-9}$G on the horizon scale, 
while 
$2.9 \times 10^{-56}$G on 1Mpc.}. 
%%%
%%%%%
We also remark that future CMB polarization experiments 
such as PLANCK~\cite{Planck-1, Planck-2}, 
QUIET~\cite{QUIET-1, Samtleben:2008rb}, 
B-Pol~\cite{B-Pol} and LiteBIRD~\cite{LiteBIRD}
can test the large-scale magnetic fields with 
the current amplitude $\sim 4 \times 10^{-11} - 10^{-10}$G~\cite{Test}. 
%%%%%

%%%%%%%%%%%%%%%%%%%
%%%  Sec. IV
%%%%%%%%%%%%%%%%%%%
\section{conclusions}

We have studied the generation of the large-scale magnetic fields 
from inflation due to the $CPT$-even dimension-six Chern-Simons-like 
effective interaction in the presence of the dynamical Kalb-Ramond and 
scalar fields. 
It has explicitly been shown that the magnetic fields on 1Mpc scale 
with the present amplitude of $\sim 10^{-9}$G can be generated 
when the number density of the fermion interacting with 
the electromagnetic field evolves in 
proportion to $a^{-m}(t)$ with $m=1$, $2$ and $3$ during inflation. 
If the large-scale magnetic fields $\sim 10^{-9}$G are generated from 
inflation, the magnetic fields observed in galaxies and clusters of galaxies 
can be explained through only adiabatic compression without 
any dynamo amplification mechanism~\cite{Turner:1987bw}.

%%%%%%%%%%%%%%%%%%%%%%%
%%% Acknowledgments %%%
%%%%%%%%%%%%%%%%%%%%%%%
%\noindent
%{\bf Acknowledgments}
%\begin{acknowledgments}
\section*{Acknowledgments}

K.B. would like to sincerely appreciate very kind and warm hospitality 
at Eurasian National University very much. 
The work by W.F.K. is supported in part by the National Science 
Council of R.O.C. under Grant number: NSC-98-2112-M-009-002-MY3, 
that by S.H.H. is supported in part by the National Science 
Council of R.O.C. under Grant number: NSC-98-2112-M-009-002-MY3 
and NSC98-2917-I-564-122, 
and that by K.B. and C.Q.G. is supported in part 
by the National Science Council of R.O.C. 
under Grant number: NSC-98-2112-M-007-008-MY3 and 
National Tsing Hua University under the Boost Program (99N2539E1).

%\end{acknowledgments}

%%%%%%%%%%%%%%%%%%%%%%%%%%%
%%%  Appendix
%%%%%%%%%%%%%%%%%%%%%%%%%%%

\appendix

\section{Asymptotic behavior of $C_+(k,\tilde{t}\,)$ and $C_-(k,\tilde{t}\,)$}
\label{A}
%\label{A1}

We start from Eq.~(\ref{eq310}).
\beq \label{A1}
C_{\pm}^{\prime\prime}(k,\tilde{t}\,)
+C_{\pm}^{\prime}(k,\tilde{t}\,)+e^{-2(\tilde{t}-\tilde{t}_1)} 
\left[1\mp J_1 e^{(m+1)(\tilde{t}-\tilde{t}_1)} \mp 
J_2 e^{(1-m)(\tilde{t}-\tilde{t}_1)} \right] C_{\pm}(k,\tilde{t}\,)=0\,.
\eeq
If we take $C_{\pm}=e^{f_{\pm}}$, $g_{\pm}=f'_{\pm}$, $J=-J_1>0$ and $m_{-}=m-1$, the field equations can be written as 
\beq \label{A2}
g'(k, \tau)+g(k,\tau)^2+g(k,\tau)\pm J e^{m_- \tau}=0
\eeq
when $\tau \equiv \tilde{t}-\tilde{t}_1 \rightarrow \infty$. It is easy to see that the $J$ term dominates at large $\tau$ for all positive $m_-$. The homogeneous solution $g_1$ to above equation is, ignoring the initial time when writing $\tau$ as $\tau - \tau_0$, 
\beq \label{A3}
g_1=\frac{d e^{-\tau}}{1-d e^{-\tau}} \rightarrow d e^{-\tau}
\eeq
at the large time limit, with $d$ a parameter to be fitted with the initial conditions. Hence the homogeneous part does not affect the asymptotic behavior in any significant way even the non-linear term $g^2$ is present. Therefore, the large time physics is controlled mainly by the algebraic equation
\beq \label{A4}
g(k,\tau)^2+g(k,\tau) \pm J e^{m_- \tau}=0.
\eeq

For $m = 1$ Eq.~(\ref{A4}) gives, the first $\pm$ sign indicating two 
different roots, the second $\mp$ sign indicating solutions to the $C_{\pm}$ 
equation, 
\beq \label{A5}
g_{\pm}(k,\tau)=\frac{-1\pm \sqrt{1\mp 4 J}}{2}.
\eeq

\begin{subequations} \label{A6}
Therefore, we have the solution
\beq \label{A6a}
C_+ \rightarrow \exp [\frac{(-1\pm \sqrt{1-4J})\tau}{2}] \rightarrow 0
\eeq
for all $J$, with $4J>0$ indicating the oscillating solutions. In addition,
\beq 
\label{A6b}
&& C_- \rightarrow \exp [(\frac{-1+ \sqrt{1+4J}}{2})\tau] \rightarrow \infty, \\
\label{A6c}
&& C_- \rightarrow \exp [(\frac{-1- \sqrt{1+4J}}{2})\tau] \rightarrow 0.
\eeq
\end{subequations}

For $m>1$, one has
\begin{subequations} \label{A7}
\beq 
\label{A7a} && C_+ \rightarrow N \exp[\pm i (2\sqrt{J}/m_-)\exp(m_- \tau /2)] \rightarrow C_0, \ \ \mbox{a constant}, \\
\label{A7b} && C_- \rightarrow N \exp[\pm (2\sqrt{J}/m_-) \exp(m_- \tau /2)] .
\eeq
\end{subequations}
It is clear that $C_+$ is an oscillatory solution and $C_-$ has two solutions, one approaching to infinity (+) and the other to zero (-). A small $-1/2$ has been ignored when solving for $g$ in above solutions. This has to do with the fact that we have ignored the homogeneous part of $g$ equation. 

\begin{center}
\begin{figure}[tbp]
\begin{tabular}{ll}
\begin{minipage}{80mm}
\begin{center}
\unitlength=1mm
\resizebox{!}{6.5cm}{
   \includegraphics{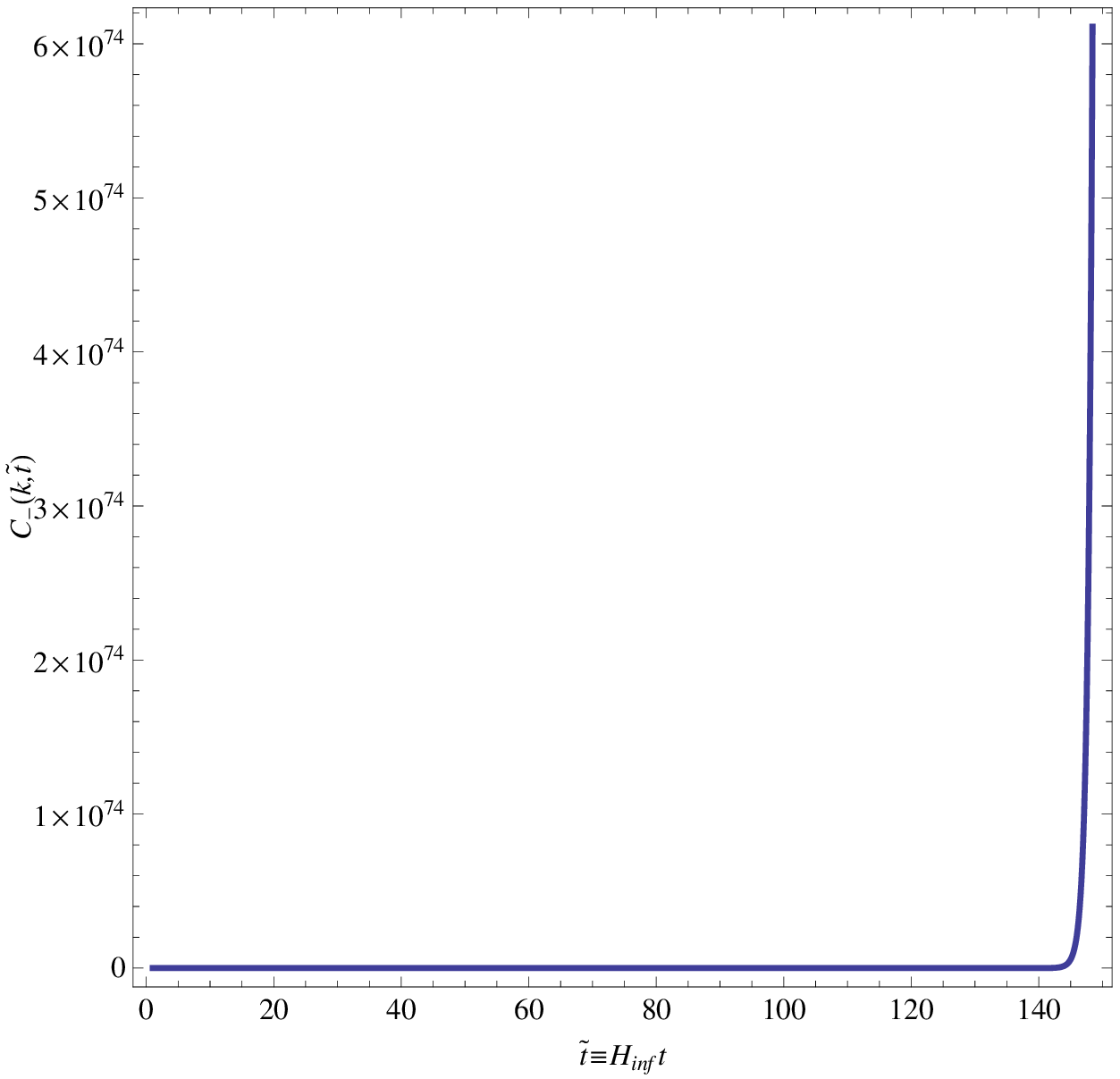}
                  }
\end{center}
\end{minipage}
&
\begin{minipage}{80mm}
\begin{center}
\unitlength=1mm
\resizebox{!}{6.5cm}{
   \includegraphics{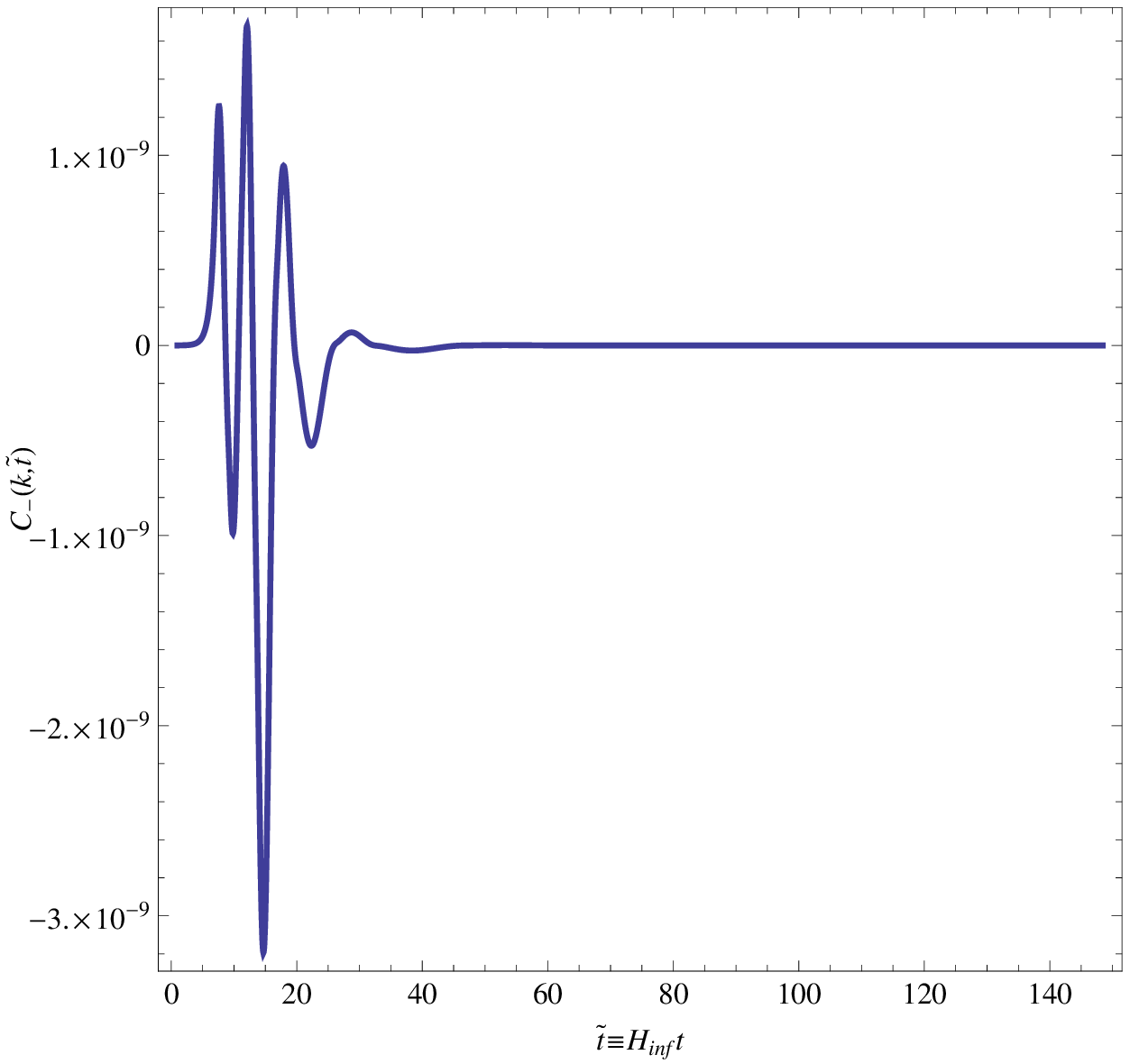}
                  }
\end{center}
\end{minipage}

\end{tabular}
\caption{$C_-(k,\tilde{t}\,)$ as functions of $\tilde{t} \equiv H_{\mathrm{inf}}t$ 
with a comoving scale $L=2\pi/k = 1\mathrm{Mpc}$ 
for $n=\bar{n}a^{-1}$ (m=1), where $\bar{n}=10^{-104}$, 
$H_{\mathrm{inf}} = 10^{10}$GeV, $V_0=10^{-47} \ \mathrm{GeV}^4$ 
and $\xi_1=\xi_2=1$. The initial conditions are $C_-(\tilde{t_1}=1)=0$ and $C'_-(\tilde{t_1}=1)=10^{-12}$ (left) and $C_-(\tilde{t_1}=1)=0$ and $C'_-(\tilde{t_1}=1)=10^{-12.1}$ (right).
}
\label{fig-4}
\end{figure}
\end{center}

\begin{center}
\begin{figure}[tbp]
\begin{tabular}{ll}
\begin{minipage}{80mm}
\begin{center}
\unitlength=1mm
\resizebox{!}{6.5cm}{
   \includegraphics{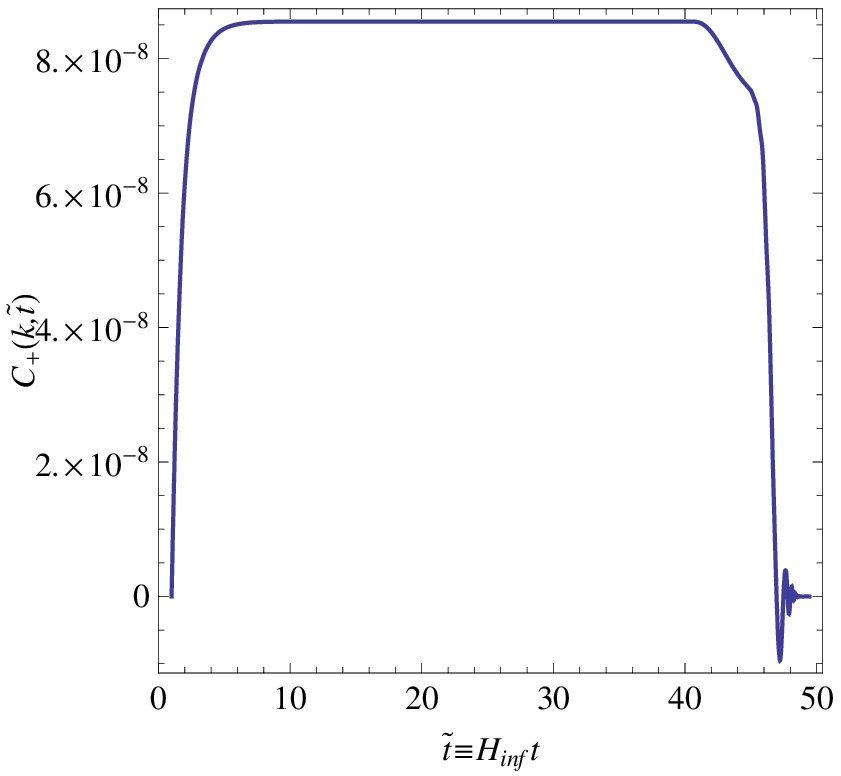}
                  }
\end{center}
\end{minipage}
&
\begin{minipage}{80mm}
\begin{center}
\unitlength=1mm
\resizebox{!}{6.5cm}{
   \includegraphics{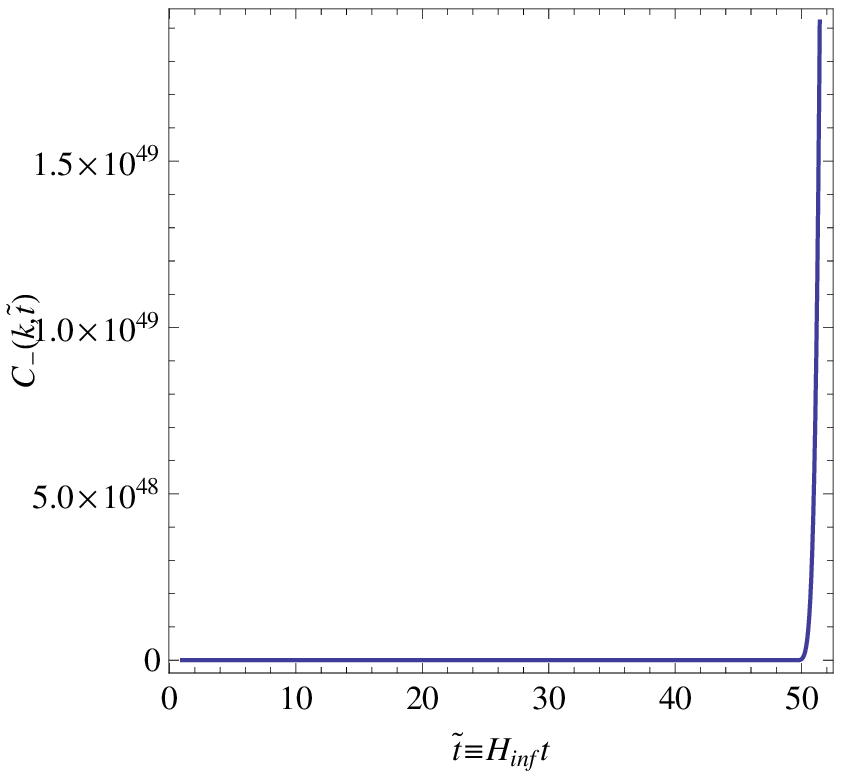}
                  }
\end{center}
\end{minipage}

\end{tabular}
\caption{$C_+(k,\tilde{t}\,)$ (left) and $C_-(k,\tilde{t}\,)$ 
(right) as functions of $\tilde{t} \equiv H_{\mathrm{inf}}t$ 
with a comoving scale $L=2\pi/k = 1\mathrm{Mpc}$ 
for $n=\bar{n}a^{-4}$ (m=4), where $\bar{n}=10^{-187}$, 
$H_{\mathrm{inf}} = 10^{10}$GeV, $V_0=10^{-47} \ \mathrm{GeV}^4$ 
and $\xi_1=\xi_2=1$. The initial conditions are $C_+(\tilde{t_1}=1)=0$ and $C'_-(\tilde{t_1}=1)=10^{-7}$ and $C_-(\tilde{t_1}=1)=0$ and $C'_-(\tilde{t_1}=1)=10^{-7}$ (right). 
}
\label{fig-5}
\end{figure}
\end{center}

\begin{center}
\begin{figure}[tbp]
\begin{tabular}{ll}
\begin{minipage}{80mm}
\begin{center}
\unitlength=1mm
\resizebox{!}{6.5cm}{
   \includegraphics{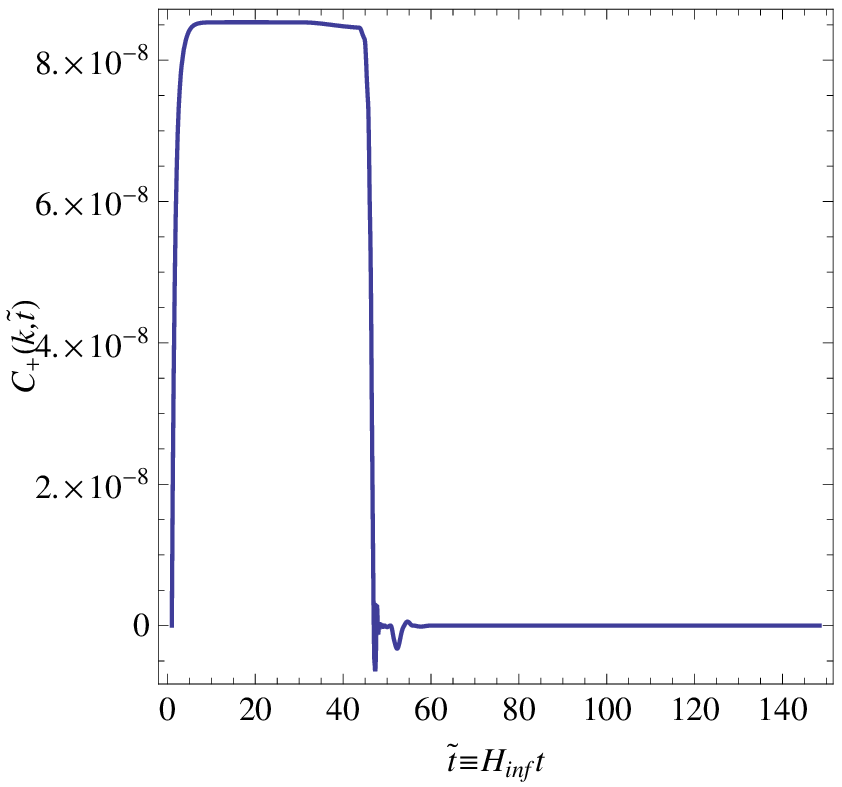}
                  }
\end{center}
\end{minipage}
&
\begin{minipage}{80mm}
\begin{center}
\unitlength=1mm
\resizebox{!}{6.5cm}{
   \includegraphics{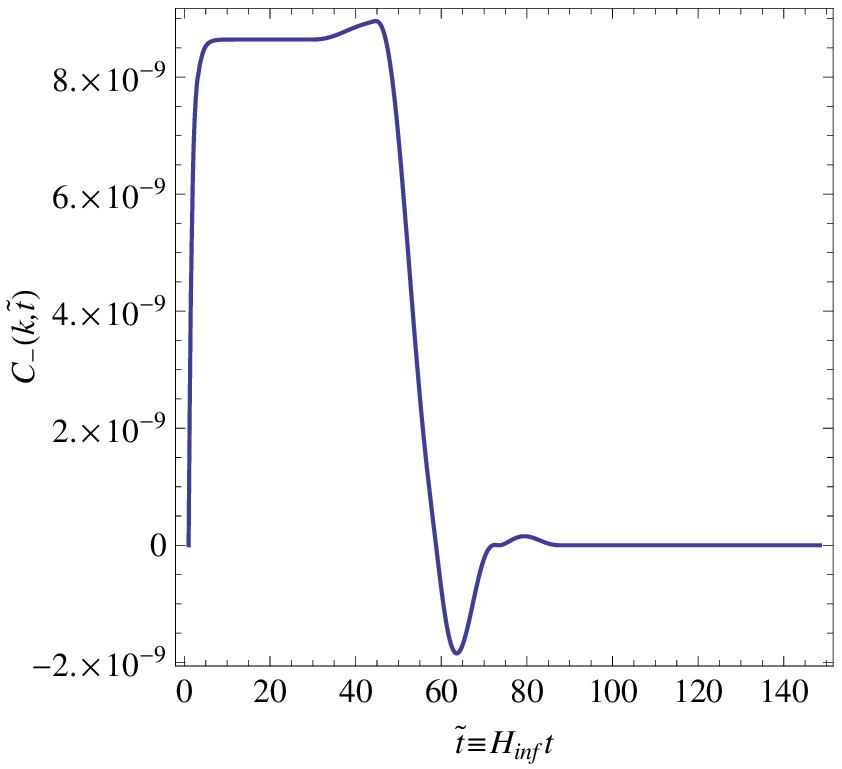}
                  }
\end{center}
\end{minipage}

\end{tabular}
\caption{Legend is the same as Fig. (\ref{fig-5}) but with initial conditions $C_+(\tilde{t_1}=1)=0$ and $C'_-(\tilde{t_1}=1)=10^{-7}$ and $C_-(\tilde{t_1}=1)=0$ and $C'_-(\tilde{t_1}=1)=10^{-8}$ (right). 
}
\label{fig-6}
\end{figure}
\end{center}

In summary, we have found there is only one solution of $C_+$ 
at the large time limit and there are, however, two sets of $C_-$ solutions for both $m=1$ and $m>1$. The numerical study shows that the behavior of $C_-$ is very sensitive to the choice of the initial condition as shown in Figs. (\ref{fig-4}), (\ref{fig-5}) and (\ref{fig-6}).

%%%%%%%%%%%%%%%%%%%%%
%%%  Figures
%%%%%%%%%%%%%%%%%%%%%
%%%%%% Fig. 4 %%%%%%%%%
%%%%%%%%%%%%%%%%%%%%%%%%

%%%%%%%%%%%%%%%%%%%%%
%%%  Figures
%%%%%%%%%%%%%%%%%%%%%
%%%%%% Fig. 5 %%%%%%%%%
%%%%%%%%%%%%%%%%%%%%%%%%

%%%%%%%%%%%%%%%%%%%%%
%%%  Figures
%%%%%%%%%%%%%%%%%%%%%
%%%%%% Fig. 6 %%%%%%%%%
%%%%%%%%%%%%%%%%%%%%%%%%

\newpage

%%%%%%%%%%%%%%%%%%%%%%%%%%%%%%%%%
%% thebibliography environment 
%%%%%%%%%%%%%%%%%%%%%%%%%%%%%%%%%


\begin{thebibliography}{99}

\bibitem{reviews}
%For reviews on cosmic magnetic fields, see 
%\cite{Kronberg:1993vk}
%\bibitem{Kronberg:1993vk}
  P.~P.~Kronberg,
  %``Extragalactic magnetic fields,''
  Rept.\ Prog.\ Phys.\  {\bf 57}, 325 (1994);\ 
  %%CITATION = RPPHA,57,325;%%
%\cite{Grasso:2000wj}
%\bibitem{Grasso:2000wj}
  D.~Grasso and H.~R.~Rubinstein,
  %``Magnetic fields in the early universe,''
  Phys.\ Rept.\  {\bf 348}, 163 (2001);\ 
%  [arXiv:astro-ph/0009061].
  %%CITATION = PRPLC,348,163;%%
%\bibitem{Carilli1}
  C.~L.~Carilli and G.~B.~Taylor, 
  Ann.\ Rev.\ Astron.\ Astrophys.\ {\bf 40}, 319 (2002);\ 
%\cite{Widrow:2002ud}
%\bibitem{Widrow:2002ud}
  L.~M.~Widrow,
  %``Origin of Galactic and Extragalactic Magnetic Fields,''
  Rev.\ Mod.\ Phys.\  {\bf 74}, 775 (2002);\ 
%  [arXiv:astro-ph/0207240].
  %%CITATION = RMPHA,74,775;%%
%\cite{Giovannini:2003yn}
%\bibitem{Giovannini:2003yn}
  M.~Giovannini,
  %``The magnetized universe,''
  Int.\ J.\ Mod.\ Phys.\  D {\bf 13}, 391 (2004);\ 
%  [arXiv:astro-ph/0312614].
  %%CITATION = IMPAE,D13,391;%%
%\cite{Giovannini:2006kg}
%\bibitem{Giovannini:2006kg}
%  M.~Giovannini,
  %``Magnetic fields, strings and cosmology,''
  Lect.\ Notes Phys.\  {\bf 737}, 863 (2008). 
%  [arXiv:astro-ph/0612378].
  %%CITATION = LNPHA,737,863;%%

%\cite{Kandus:2010nw}
\bibitem{Kandus:2010nw}
  A.~Kandus, K.~E.~Kunze and C.~G.~Tsagas,
  %``Primordial magnetogenesis,''
  Phys.\ Rept.\  {\bf 505}, 1 (2011). 
%  [arXiv:1007.3891 [astro-ph.CO]].
  %%CITATION = PRPLC,505,1;%%

\bibitem{EParker}
E.~N.~Parker,
Astrophys.\ J.\ {\bf 163}, 255 (1971);\
\textit{Cosmical Magnetic Fields}
(Clarendon, Oxford, England, 1979);\
Ya.~B.~Zel'dovich, A.~A.~Ruzmaikin, and D.~D.~Sokoloff,
\textit{Magnetic Fields in Astrophysics}
(Gordon and Breach, New York, 1983).

\bibitem{Biermann1}
L.~Biermann and A.~Schl\"{u}ter,
Phys.\ Rev.\ {\bf 82}, 863 (1951). 

\bibitem{PI}
%\cite{Weibel:1959zz}
%\bibitem{Weibel:1959zz}
  E.~S.~Weibel,
  %``Spontaneously Growing Transverse Waves In A Plasma Due To An Anisotropic
  %Veloc Ity Distribution,''
  Phys.\ Rev.\ Lett.\  {\bf 2}, 83 (1959).

\bibitem{PT}
%\bibitem{Quashnock}
J.~M.~Quashnock, A.~Loeb, and D.~N.~Spergel,
Astrophys.\ J.\ {\bf 344}, L49 (1989);\ 
%\cite{Baym:1995fk}
%\bibitem{Baym:1995fk}
  G.~Baym, D.~Bodeker and L.~D.~McLerran,
  %``Magnetic fields produced by phase transition bubbles in the electroweak
  %phase transition,''
  Phys.\ Rev.\  D {\bf 53}, 662 (1996);\ 
%  [arXiv:hep-ph/9507429].
  %%CITATION = PHRVA,D53,662;%%
%\cite{Boyanovsky:2002wa}
%\bibitem{Boyanovsky:2002wa}
  D.~Boyanovsky, H.~J.~de Vega and M.~Simionato,
  %``Large scale magnetogenesis from a non-equilibrium phase transition in the
  %radiation dominated era,''
%  Phys.\ Rev.\  D {\bf 67}, 123505 (2003);\ 
\textit{ibid}.\ {\bf 67}, 123505 (2003);\ 
%  [arXiv:hep-ph/0211022];\
  %%CITATION = PHRVA,D67,123505;%%
%\cite{Boyanovsky:2002kq}
%\bibitem{Boyanovsky:2002kq}
%  D.~Boyanovsky, M.~Simionato and H.~J.~de Vega,
  %``Magnetic field generation from non-equilibrium phase transitions,''
%  Phys.\ Rev.\  D {\bf 67}, 023502 (2003)
{\bf 67}, 023502 (2003);\ 
%  [arXiv:hep-ph/0208272].
  %%CITATION = PHRVA,D67,023502;%%
%\cite{Durrer:2003ja}
%\bibitem{Durrer:2003ja}
  R.~Durrer and C.~Caprini,
  %``Primordial Magnetic Fields and Causality,''
  JCAP {\bf 0311}, 010 (2003);\ 
%  [arXiv:astro-ph/0305059].
  %%CITATION = JCAPA,0311,010;%%
%
%\cite{Kahniashvili:2009qi}
%\bibitem{Kahniashvili:2009qi}
  T.~Kahniashvili, A.~G.~Tevzadze and B.~Ratra,
  %``Phase Transition Generated Cosmological Magnetic Field at Large Scales,''
  Astrophys.\ J.\  {\bf 726}, 78 (2011). 
%  [arXiv:0907.0197 [astro-ph.CO]].
  %%CITATION = ASJOA,726,78;%%
%

\bibitem{DP}
%\cite{Berezhiani:2003ik}
%\bibitem{Berezhiani:2003ik}
  Z.~Berezhiani and A.~D.~Dolgov,
  %``Generation of large scale magnetic fields at recombination epoch,''
  Astropart.\ Phys.\  {\bf 21}, 59 (2004);\ 
%  [arXiv:astro-ph/0305595].
  %%CITATION = APHYE,21,59;%%
%\cite{Matarrese:2004kq}
%\bibitem{Matarrese:2004kq}
  S.~Matarrese, S.~Mollerach, A.~Notari and A.~Riotto, 
%S.~Matarrese {\it et al.}, 
  %``Large-scale magnetic fields from density perturbations,''
  Phys.\ Rev.\  D {\bf 71}, 043502 (2005);\ 
%  [arXiv:astro-ph/0410687].
  %%CITATION = PHRVA,D71,043502;%%
%\cite{Takahashi:2005nd}
%\bibitem{Takahashi:2005nd}
  K.~Takahashi, K.~Ichiki, H.~Ohno and H.~Hanayama, 
%K.~Takahashi {\it et al.}, 
  %``Magnetic field generation from cosmological perturbations,''
  Phys.\ Rev.\ Lett.\  {\bf 95}, 121301 (2005);\ 
%  [arXiv:astro-ph/0502283].
  %%CITATION = PRLTA,95,121301;%%
%\bibitem{Ichiki1}
%K.~Ichiki \textit{et al}., 
K.~Ichiki, K.~Takahashi, H.~Ohno, H.~Hanayama, and N.~Sugiyama,
Science {\bf 311}, 827 (2006);\ 
%\cite{Siegel:2006px}
%\bibitem{Siegel:2006px}
  E.~R.~Siegel and J.~N.~Fry, 
  %``Cosmological Structure Formation Creates Large-Scale Magnetic Fields,''
  Astrophys.\ J.\  {\bf 651}, 627 (2006);\ 
%  [arXiv:astro-ph/0604526].
  %%CITATION = ASJOA,651,627;%%
%\cite{Kobayashi:2007wd}
%\bibitem{Kobayashi:2007wd}
  T.~Kobayashi, R.~Maartens, T.~Shiromizu and K.~Takahashi, 
%T.~Kobayashi {\it et al.}, 
  %``Cosmological magnetic fields from nonlinear effects,''
  Phys.\ Rev.\  D {\bf 75}, 103501 (2007);\ 
%  [arXiv:astro-ph/0701596].
  %%CITATION = PHRVA,D75,103501;%%
%
%%%%%
K.~E.~Kunze,
%``Primordial magnetic fields and nonlinear electrodynamics,''
%Phys.\ Rev.\  D {\bf 77}, 023530 (2008);
\textit{ibid}. {\bf 77}, 023530 (2008);\ 
L.~Campanelli, P.~Cea, G.~L.~Fogli and L.~Tedesco,
%``Inflation-Produced Magnetic Fields in Nonlinear Electrodynamics,''
%  Phys.\ Rev.\  D {\bf 77}, 043001 (2008)
\textit{ibid}. {\bf 77}, 043001 (2008);\ 
%
%\cite{Maeda:2008dv}
%\bibitem{Maeda:2008dv}
  S.~Maeda, S.~Kitagawa, T.~Kobayashi and T.~Shiromizu,
  %``Primordial magnetic fields from second-order cosmological
  %perturbations:Tight coupling approximation,''
  Class.\ Quant.\ Grav.\  {\bf 26}, 135014 (2009);\ 
%  [arXiv:0805.0169 [astro-ph]].
  %%CITATION = CQGRD,26,135014;%%
%
%\cite{Fenu:2010kh}
%\bibitem{Fenu:2010kh}
  E.~Fenu, C.~Pitrou and R.~Maartens,
  %``The seed magnetic field generated during recombination,''
  arXiv:1012.2958 [astro-ph.CO].
  %%CITATION = ARXIV:1012.2958;%%
%
%%%%%

%\cite{Turner:1987bw}
\bibitem{Turner:1987bw}
  M.~S.~Turner and L.~M.~Widrow,
  %``Inflation Produced, Large Scale Magnetic Fields,''
  Phys.\ Rev.\  D {\bf 37}, 2743 (1988).
  %%CITATION = PHRVA,D37,2743;%%

\bibitem{Backreaction}
%
%\cite{Demozzi:2009fu}
%\bibitem{Demozzi:2009fu}
  V.~Demozzi, V.~Mukhanov and H.~Rubinstein,
  %``Magnetic fields from inflation?,''
  JCAP {\bf 0908}, 025 (2009);\ 
%  [arXiv:0907.1030 [astro-ph.CO]].
  %%CITATION = JCAPA,0908,025;%%
%
%\cite{Kanno:2009ei}
%\bibitem{Kanno:2009ei}
  S.~Kanno, J.~Soda and M.~a.~Watanabe,
  %``Cosmological Magnetic Fields from Inflation and Backreaction,''
%  JCAP {\bf 0912}, 009 (2009)
\textit{ibid}. {\bf 0912}, 009 (2009). 
%  [arXiv:0908.3509 [astro-ph.CO]].
  %%CITATION = JCAPA,0912,009;%%
%

%\cite{Parker:1968mv}
\bibitem{Parker:1968mv}
  L.~Parker,
  %``Particle creation in expanding universes,''
  Phys.\ Rev.\ Lett.\  {\bf 21}, 562 (1968).
  %%CITATION = PRLTA,21,562;%%


%%%%%
%\cite{Maroto:2000zu}
\bibitem{Maroto:2000zu}
  A.~L.~Maroto,
  %``Primordial magnetic fields from metric perturbations,''
  Phys.\ Rev.\  D {\bf 64}, 083006 (2001). 
%  [arXiv:hep-ph/0008288].
  %%CITATION = PHRVA,D64,083006;%%
%%%%%

\bibitem{NZC}
%
%\cite{Tsagas:2004kv}
%\bibitem{Tsagas:2004kv}
  C.~G.~Tsagas,
  %``Electromagnetic fields in curved spacetimes,''
  Class.\ Quant.\ Grav.\  {\bf 22}, 393 (2005);\ 
%  [arXiv:gr-qc/0407080].
  %%CITATION = CQGRD,22,393;%%
%
%\cite{Tsagas:2005nn}
%\bibitem{Tsagas:2005nn}
  C.~G.~Tsagas and A.~Kandus, 
  %``Superadiabatic-type magnetic amplification in conventional cosmology,''
  Phys.\ Rev.\  D {\bf 71}, 123506 (2005);\ 
%  [arXiv:astro-ph/0504089].
  %%CITATION = PHRVA,D71,123506;%%
%\cite{Barrow:2008jp}
%\bibitem{Barrow:2008jp}
  J.~D.~Barrow and C.~G.~Tsagas,
  %``Slow decay of magnetic fields in open Friedmann universes,''
%  Phys.\ Rev.\  D {\bf 77}, 107302 (2008)
\textit{ibid}.\ {\bf 77}, 107302 (2008)
  [Erratum-ibid.\  D {\bf 77}, 109904 (2008)];\ 
%  [arXiv:0803.0660 [astro-ph]].
  %%CITATION = PHRVA,D77,107302;%%
%
%\cite{Barrow:2011ic}
%\bibitem{Barrow:2011ic}
%  J.~D.~Barrow and C.~G.~Tsagas,
  %``Cosmological magnetic field survival,''
  Mon.\ Not.\ Roy.\ Astron.\ Soc.\  {\bf 414}, 512 (2011). 
%  [arXiv:1101.2390 [astro-ph.CO]].
  %%CITATION = MNRAA,414,512;%%
%

%%%%%
\bibitem{NGC}
F.~D.~Mazzitelli and F.~M.~Spedalieri,
%``Scalar electrodynamics and primordial magnetic fields,''
Phys.\ Rev.\  D {\bf 52}, 6694 (1995); 
G.~Lambiase and A.~R.~Prasanna,
\textit{ibid}.\ {\bf 70}, 063502 (2004); 
%
%\cite{Bamba:2008ja}
%\bibitem{Bamba:2008ja}
  K.~Bamba and S.~D.~Odintsov,
  %``Inflation and late-time cosmic acceleration in non-minimal Maxwell-$F(R)$
  %gravity and the generation of large-scale magnetic fields,''
  JCAP {\bf 0804}, 024 (2008); 
%\textit{ibid}.\  {\bf 0804}, 024 (2008). 
%  [arXiv:0801.0954 [astro-ph]].
  %%CITATION = JCAPA,0804,024;%%
%
%\cite{Campanelli:2008qp}
%\bibitem{Campanelli:2008qp}
  L.~Campanelli, P.~Cea, G.~L.~Fogli and L.~Tedesco,
  %``Inflation-Produced Magnetic Fields in R^n F^2 and I F^2 models,''
  Phys.\ Rev.\  D {\bf 77}, 123002 (2008);\ 
%\textit{ibid}.\ {\bf 77}, 123002 (2008); 
%  [arXiv:0802.2630 [astro-ph]].
  %%CITATION = PHRVA,D77,123002;%%
%
%\cite{Lambiase:2008zz}
%\bibitem{Lambiase:2008zz}
  G.~Lambiase, S.~Mohanty and G.~Scarpetta,
  %``Magnetic field amplification in f(R) theories of gravity,''
  JCAP {\bf 0807}, 019 (2008);\ 
  %%CITATION = JCAPA,0807,019;%%
%
%\cite{Kunze:2009bs}
%\bibitem{Kunze:2009bs}
  K.~E.~Kunze,
  %``Large scale magnetic fields from gravitationally coupled 
  %electrodynamics,''
  Phys.\ Rev.\  D {\bf 81}, 043526 (2010);\ 
%\textit{ibid}.\ {\bf 81}, 043526 (2010). 
%  [arXiv:0911.1101 [astro-ph.CO]].
  %%CITATION = PHRVA,D81,043526;%%
%
%\cite{Jimenez:2010uh}
%\bibitem{Jimenez:2010uh}
  J.~B.~Jimenez and A.~L.~Maroto,
  %``Dark energy, non-minimal couplings and the origin of cosmic magnetic
  %fields,''
  JCAP {\bf 1012}, 025 (2010). 
%  [arXiv:1010.4513 [astro-ph.CO]].
  %%CITATION = JCAPA,1012,025;%%
%
%%%%%

%%%%%
%\cite{Tsagas:2009cr}
\bibitem{Tsagas:2009cr}
  C.~G.~Tsagas,
  %``Gravito-magnetic amplification in cosmology,''
  Phys.\ Rev.\  D {\bf 81}, 043501 (2010). 
%  [arXiv:0912.2749 [astro-ph.CO]].
  %%CITATION = PHRVA,D81,043501;%%
%%%%%

%\cite{Ratra:1991bn}
\bibitem{Ratra:1991bn}
  B.~Ratra,
  %``Cosmological 'seed' magnetic field from inflation,''
  Astrophys.\ J.\  {\bf 391}, L1 (1992).
  %%CITATION = ASJOA,391,L1;%%

\bibitem{DE}
%\cite{Lemoine:1995vj}
%\bibitem{Lemoine:1995vj}
  D.~Lemoine and M.~Lemoine,
  %``Primordial magnetic fields in string cosmology,''
  Phys.\ Rev.\  D {\bf 52}, 1955 (1995);\ 
  %%CITATION = PHRVA,D52,1955;%%
%\cite{Gasperini:1995dh}
%\bibitem{Gasperini:1995dh}
  M.~Gasperini, M.~Giovannini and G.~Veneziano,
  %``Primordial magnetic fields from string cosmology,''
  Phys.\ Rev.\ Lett.\  {\bf 75}, 3796 (1995);\ 
%  [arXiv:hep-th/9504083].
  %%CITATION = PRLTA,75,3796;%%
%\cite{Bamba:2003av}
%\bibitem{Bamba:2003av}
  K.~Bamba and J.~Yokoyama,
  %``Large-scale magnetic fields from inflation in dilaton electromagnetism,''
  Phys.\ Rev.\  D {\bf 69}, 043507 (2004);\ 
%  [arXiv:astro-ph/0310824];\
  %%CITATION = PHRVA,D69,043507;%%
%\cite{Bamba:2004cu}
%\bibitem{Bamba:2004cu}
%  K.~Bamba and J.~Yokoyama,
  %``Large-scale magnetic fields from dilaton inflation in noncommutative
  %spacetime,''
%  Phys.\ Rev.\  D {\bf 70}, 083508 (2004)
{\bf 70}, 083508 (2004);\ 
%  [arXiv:hep-ph/0409237].
  %%CITATION = PHRVA,D70,083508;%%
%\cite{Salim:2006nw}
%\bibitem{Salim:2006nw}
  J.~M.~Salim, N.~Souza, S.~E.~Perez Bergliaffa and T.~Prokopec, 
%J.~M.~Salim {\it et al.}, 
  %``Creation of cosmological magnetic fields in a bouncing cosmology,''
  JCAP {\bf 0704}, 011 (2007);\ 
%  [arXiv:astro-ph/0612281].
  %%CITATION = JCAPA,0704,011;%%
% 
%\cite{AkhtariZavareh:2007zz}
%\bibitem{AkhtariZavareh:2007zz}
  A.~Akhtari-Zavareh, A.~Hojati and B.~Mirza,
  %``Generation of large scale magnetic fields by coupling to curvature and
  %dilaton field,''
  Prog.\ Theor.\ Phys.\  {\bf 117}, 803 (2007);\ 
%  [arXiv:0707.3493 [astro-ph]].
  %%CITATION = PTPKA,117,803;%%
%
%\cite{Martin:2007ue}
%\bibitem{Martin:2007ue}
  J.~Martin and J.~Yokoyama, 
  %``Generation of Large-Scale Magnetic Fields in Single-Field Inflation,''
  JCAP {\bf 0801}, 025 (2008);\ 
%\textit{ibid}.\ {\bf 0801}, 025 (2008). 
%  [arXiv:0711.4307 [astro-ph]].
  %%CITATION = JCAPA,0801,025;%%
%
%\cite{Das:2010ywa}
%\bibitem{Das:2010ywa}
  M.~Das and S.~Mohanty,
  %``Magnetic field generation in Higgs inflation model,''
  arXiv:1004.1927 [astro-ph.CO].
  %%CITATION = ARXIV:1004.1927;%%
%

\bibitem{C-S}
%\cite{Giovannini:2001nh}
%\bibitem{Giovannini:2001nh}
  M.~Giovannini,
  %``On the variation of the gauge couplings during inflation,''
  Phys.\ Rev.\  D {\bf 64}, 061301 (2001);\ 
%\textit{ibid}. {\bf 64}, 061301 (2001);\
%  [arXiv:astro-ph/0104290];\
  %%CITATION = PHRVA,D64,061301;%%
%\cite{Bertolami:2005np}
%\bibitem{Bertolami:2005np}
  O.~Bertolami and R.~Monteiro, 
  %``Varying electromagnetic coupling and primordial magnetic fields,''
%  Phys.\ Rev.\  D {\bf 71}, 123525 (2005);\ 
\textit{ibid}.\ {\bf 71}, 123525 (2005);\ 
%  [arXiv:astro-ph/0504211].
  %%CITATION = PHRVA,D71,123525;%%
%
%\cite{Giovannini:2007rh}
%\bibitem{Giovannini:2007rh}
  M.~Giovannini,
  %``Magnetogenesis, spectator fields and CMB signatures,''
  Phys.\ Lett.\  B {\bf 659}, 661 (2008);\ 
%  [arXiv:0711.3273 [astro-ph]].
  %%CITATION = PHLTA,B659,661;%%
%
%\cite{Seery:2008ms}
%\bibitem{Seery:2008ms}
  D.~Seery,
  %``Magnetogenesis and the primordial non-gaussianity,''
  JCAP {\bf 0908}, 018 (2009). 
%  [arXiv:0810.1617 [astro-ph]].
  %%CITATION = JCAPA,0908,018;%%
%

\bibitem{pseudoscalar}
%\bibitem{scalar-field electromagnetism}
  %\cite{Garretson:1992vt}
%\bibitem{Garretson:1992vt}
  W.~D.~Garretson, G.~B.~Field and S.~M.~Carroll,
  %``Primordial magnetic fields from pseudoGoldstone bosons,''
  Phys.\ Rev.\  D {\bf 46}, 5346 (1992);\ 
%  [arXiv:hep-ph/9209238];\
  %%CITATION = PHRVA,D46,5346;%%
%\cite{Field:1998hi}
%\bibitem{Field:1998hi}
  G.~B.~Field and S.~M.~Carroll,
  %``Cosmological magnetic fields from primordial helicity,''
%  Phys.\ Rev.\  D {\bf 62}, 103008 (2000).
\textit{ibid}.\ {\bf 62}, 103008 (2000);\ 
%  [arXiv:astro-ph/9811206];\
  %%CITATION = PHRVA,D62,103008;%%
%\cite{Anber:2006xt}
%\bibitem{Anber:2006xt}
  M.~M.~Anber and L.~Sorbo, 
  %``N-flationary magnetic fields,''
  JCAP {\bf 0610}, 018 (2006);\ 
%  [arXiv:astro-ph/0606534].
  %%CITATION = JCAPA,0610,018;%%
%\cite{Campanelli:2008kh}
%\bibitem{Campanelli:2008kh}
  L.~Campanelli,
  %``Helical Magnetic Fields from Inflation,''
  Int.\ J.\ Mod.\ Phys.\  D {\bf 18}, 1395 (2009);\ 
%  [arXiv:0805.0575 [astro-ph]].
  %%CITATION = IMPAE,D18,1395;%%
%
%\cite{Andrianov:2008hn}
%\bibitem{Andrianov:2008hn}
  A.~A.~Andrianov, F.~Cannata, A.~Y.~Kamenshchik and D.~Regoli,
  %``Two-field cosmological models and large-scale cosmic magnetic fields,''
  JCAP {\bf 0810}, 019 (2008). 
%  [arXiv:0806.1844 [hep-th]].
  %%CITATION = JCAPA,0810,019;%%
%

%%%%%%%%
\bibitem{Axions-Giovannini-B-BGH}
%
%\cite{Giovannini:1999by}
%\bibitem{Giovannini:1999by}
  M.~Giovannini,
  %``Hypermagnetic knots, Chern-Simons waves and the baryon asymmetry,''
  Phys.\ Rev.\  D {\bf 61}, 063502 (2000);\ 
%{\bf 61}, 063502 (2000). 
%  [arXiv:hep-ph/9906241].
  %%CITATION = PHRVA,D61,063502;%%
%
%\cite{Giovannini:1999wv}
%\bibitem{Giovannini:1999wv}
%  M.~Giovannini,
  %``Primordial hypermagnetic knots,''
%  Phys.\ Rev.\  D {\bf 61}, 063004 (2000)
{\bf 61}, 063004 (2000);\ 
%\textit{ibid}. {\bf 61}, 063004 (2000). 
%  [arXiv:hep-ph/9905358].
  %%CITATION = PHRVA,D61,063004;%%
%
%
%\bibitem{B-BGH-Baryon-asymmetry-Hypermagnetic-Baryogenesis}
%
%\cite{Bamba:2006km}
%\bibitem{Bamba:2006km}
  K.~Bamba,
  %``Baryon asymmetry from hypermagnetic helicity in dilaton hypercharge
  %electromagnetism,''
%  Phys.\ Rev.\  D {\bf 74}, 123504 (2006);\ 
\textit{ibid}. {\bf 74}, 123504 (2006);\ 
%  [arXiv:hep-ph/0611152].
  %%CITATION = PHRVA,D74,123504;%%
%
%\cite{Bamba:2007hf}
%\bibitem{Bamba:2007hf}
  K.~Bamba, C.~Q.~Geng and S.~H.~Ho,
  %``Hypermagnetic Baryogenesis,''
  Phys.\ Lett.\  B {\bf 664}, 154 (2008). 
%  [arXiv:0712.1523 [hep-ph]].
  %%CITATION = PHLTA,B664,154;%%
%
%%%%%%%%

\bibitem{charged scalar}
%\cite{Calzetta:1997ku}
%\bibitem{Calzetta:1997ku}
  E.~A.~Calzetta, A.~Kandus and F.~D.~Mazzitelli,
  %``Primordial magnetic fields induced by cosmological particle creation,''
  Phys.\ Rev.\  D {\bf 57}, 7139 (1998);\ 
%  [arXiv:astro-ph/9707220];\
  %%CITATION = PHRVA,D57,7139;%%
%\cite{Kandus:1999st}
%\bibitem{Kandus:1999st}
  A.~Kandus, E.~A.~Calzetta, F.~D.~Mazzitelli and C.~E.~M.~Wagner, 
%A.~Kandus {\it et al.}, 
  %``Cosmological magnetic fields from gauge mediated supersymmetry-breaking
  %models,''
  Phys.\ Lett.\  B {\bf 472}, 287 (2000);\ 
%  [arXiv:hep-ph/9908524];\
  %%CITATION = PHLTA,B472,287;%%
%\cite{Giovannini:2000dj}
%\bibitem{Giovannini:2000dj}
  M.~Giovannini and M.~E.~Shaposhnikov,
  %``Primordial magnetic fields from inflation?,''
  Phys.\ Rev.\  D {\bf 62}, 103512 (2000). 
%  [arXiv:hep-ph/0004269].
  %%CITATION = PHRVA,D62,103512;%%

\bibitem{ScalarED}
%\cite{Davis:2000zp}
%\bibitem{Davis:2000zp}
  A.~C.~Davis, K.~Dimopoulos, T.~Prokopec and O.~Tornkvist, 
%A.~C.~Davis {\it et al.}, 
  %``Primordial spectrum of gauge fields from inflation,''
  Phys.\ Lett.\  B {\bf 501}, 165 (2001) 
  [Phys.\ Rev.\ Focus {\bf 10}, STORY9 (2002)];\ 
%  [arXiv:astro-ph/0007214];\ 
  %%CITATION = 00627,10,STORY9;%%
%
%\cite{Dimopoulos:2001wx}
%\bibitem{Dimopoulos:2001wx}
  K.~Dimopoulos, T.~Prokopec, O.~Tornkvist and A.~C.~Davis, 
%K.~Dimopoulos {\it et al.}, 
  %``Natural magnetogenesis from inflation,''
  Phys.\ Rev.\  D {\bf 65}, 063505 (2002);\ 
%  [arXiv:astro-ph/0108093];\
  %%CITATION = PHRVA,D65,063505;%%
%\cite{Prokopec:2002jn}
%\bibitem{Prokopec:2002jn}
  T.~Prokopec, O.~Tornkvist and R.~P.~Woodard,
  %``Photon mass from inflation,''
  Phys.\ Rev.\ Lett.\  {\bf 89}, 101301 (2002);\ 
%  [arXiv:astro-ph/0205331].
  %%CITATION = PRLTA,89,101301;%%
%
%%%%%
%\cite{Emami:2009vd}
%\bibitem{Emami:2009vd}
  R.~Emami, H.~Firouzjahi and M.~S.~Movahed,
  %``Inflation from Charged Scalar and Primordial Magnetic Fields?,''
  Phys.\ Rev.\  D {\bf 81}, 083526 (2010).
%\textit{ibid}.\ {\bf 81}, 083526 (2010). 
%  [arXiv:0908.4161 [hep-th]].
  %%CITATION = PHRVA,D81,083526;%%
%%%%%

\bibitem{ScalarED-2}
%\cite{Finelli:2000sh}
%\bibitem{Finelli:2000sh}
  F.~Finelli and A.~Gruppuso,
  %``Resonant amplification of gauge fields in expanding universe,''
  Phys.\ Lett.\  B {\bf 502}, 216 (2001);\ 
%  [arXiv:hep-ph/0001231].
  %%CITATION = PHLTA,B502,216;%%
%\cite{Bassett:2000aw}
%\bibitem{Bassett:2000aw}
  B.~A.~Bassett, G.~Pollifrone, S.~Tsujikawa and F.~Viniegra, 
%B.~A.~Bassett {\it et al.}, 
  %``Preheating as cosmic magnetic dynamo,''
  Phys.\ Rev.\  D {\bf 63}, 103515 (2001). 
%  [arXiv:astro-ph/0010628].
  %%CITATION = PHRVA,D63,103515;%%

%\cite{Enqvist:2004yy}
\bibitem{Enqvist:2004yy}
  K.~Enqvist, A.~Jokinen and A.~Mazumdar, 
  %``Seed perturbations for primordial magnetic fields from MSSM flat
  %directions,''
  JCAP {\bf 0411}, 001 (2004). 
%  [arXiv:hep-ph/0404269].
  %%CITATION = JCAPA,0411,001;%%

\bibitem{DBI}
%\cite{Garousi:2004hy}
%\bibitem{Garousi:2004hy}
  M.~R.~Garousi, M.~Sami and S.~Tsujikawa, 
  %``Generation of electromagnetic fields in string cosmology with a massive
  %scalar field on the anti D-brane,''
  Phys.\ Lett.\  B {\bf 606}, 1 (2005);\ 
%  [arXiv:hep-th/0405012].
  %%CITATION = PHLTA,B606,1;%%
%\cite{Ganjali:2005sr}
%\bibitem{Ganjali:2005sr}
  M.~A.~Ganjali, 
  %``DBI with primordial magnetic field in the sky,''
  JHEP {\bf 0509}, 004 (2005);\ 
%  [arXiv:hep-th/0509032].
  %%CITATION = JHEPA,0509,004;%%
%\cite{Bamba:2008my}
%\bibitem{Bamba:2008my}
  K.~Bamba, N.~Ohta and S.~Tsujikawa, 
  %``Generic estimates for magnetic fields generated during inflation including
  %Dirac-Born-Infeld theories,''
  Phys.\ Rev.\  D {\bf 78}, 043524 (2008).
%  [arXiv:0805.3862 [astro-ph]].
  %%CITATION = PHRVA,D78,043524;%%

%%%%%
%%%%%
\bibitem{BS-B}
%
%\cite{Bamba:2006ga}
%\bibitem{Bamba:2006ga}
  K.~Bamba and M.~Sasaki,
  %``Large-scale magnetic fields in the inflationary universe,''
  JCAP {\bf 0702}, 030 (2007);\ 
%  [arXiv:astro-ph/0611701].
  %%CITATION = JCAPA,0702,030;%%
%
%\cite{Bamba:2007sx}
%\bibitem{Bamba:2007sx}
  K.~Bamba,
  %``The interrelation between the generation of large-scale electric 
  %fields and 
  %that of large-scale magnetic fields during inflation,''
%  JCAP {\bf 0710}, 015 (2007). 
\textit{ibid}.\  {\bf 0710}, 015 (2007). 
%\textit{ibid}.\  {\bf 0710}, 015 (2007).
%  [arXiv:0710.1906 [astro-ph]].
  %%CITATION = JCAPA,0710,015;%%
%%%%%
%%%%%

%%%%%
\bibitem{nonlinear-electrodynamics}
%
%\cite{Kunze:2007ph}
%\bibitem{Kunze:2007ph}
  K.~E.~Kunze,
  %``Primordial magnetic fields and nonlinear electrodynamics,''
  Phys.\ Rev.\  D {\bf 77}, 023530 (2008);\ 
%  [arXiv:0710.2435 [astro-ph]].
  %%CITATION = PHRVA,D77,023530;%%
%
%\cite{MosqueraCuesta:2009tf}
%\bibitem{MosqueraCuesta:2009tf}
  H.~J.~Mosquera Cuesta and G.~Lambiase,
  %``Primordial magnetic fields and gravitational baryogenesis in nonlinear
  %electrodynamics,''
%  Phys.\ Rev.\  D {\bf 80}, 023013 (2009)
\textit{ibid}.\  {\bf 80}, 023013 (2009). 
%  [arXiv:0907.3678 [astro-ph.CO]].
  %%CITATION = PHRVA,D80,023013;%%
%
%%%%%

%%%%%
%\cite{Gasperini:2000tw}
\bibitem{Gasperini:2000tw}
  M.~Gasperini,
  %``A new mechanism for the generation of primordial seeds for the cosmic
  %magnetic fields,''
  Phys.\ Rev.\  D {\bf 63}, 047301 (2001).
%  [arXiv:astro-ph/0009476].
  %%CITATION = PHRVA,D63,047301;%%
%%%%%

%%%%%
\bibitem{Gravitoelectromagnetic}
%
%\cite{Membiela:2008zb}
%\bibitem{Membiela:2008zb}
  F.~A.~Membiela and M.~Bellini,
  %``Primordial large-scale electromagnetic fields from Gravitoelectromagnetic
  %Inflation,''
  Phys.\ Lett.\  B {\bf 674}, 152 (2009);\ 
%  [arXiv:0811.0993 [gr-qc]].
  %%CITATION = PHLTA,B674,152;%%
%
%\cite{Membiela:2010rv}
%\bibitem{Membiela:2010rv}
%  F.~A.~Membiela and M.~Bellini,
  %``Coupled inflaton and electromagnetic fields from Gravitoelectromagnetic
  %Inflation with Lorentz and Feynman gauges,''
  JCAP {\bf 1010}, 001 (2010). 
%  [arXiv:1003.4175 [astro-ph.CO]].
  %%CITATION = JCAPA,1010,001;%%
%
%%%%%

%\cite{Dolgov:1993vg}
\bibitem{Dolgov:1993vg}
  A.~Dolgov,
  %``Breaking Of Conformal Invariance And Electromagnetic Field Generation In
  %The Universe,''
  Phys.\ Rev.\  D {\bf 48}, 2499 (1993). 
%  [arXiv:hep-ph/9301280].
  %%CITATION = PHRVA,D48,2499;%%

%\cite{Bertolami:1998dn}
\bibitem{Bertolami:1998dn}
  O.~Bertolami and D.~F.~Mota,
  %``Primordial magnetic fields via spontaneous breaking of Lorentz
  %invariance,''
  Phys.\ Lett.\  B {\bf 455}, 96 (1999). 
%  [arXiv:gr-qc/9811087].
  %%CITATION = PHLTA,B455,96;%%

%%%%%
\bibitem{L-V-T}
%
%\cite{Campanelli:2008xs}
%\bibitem{Campanelli:2008xs}
  L.~Campanelli and P.~Cea,
  %``Maxwell-Kosteleck\'y Electromagnetism and Cosmic Magnetization,''
  arXiv:0812.3745 [astro-ph];\ 
  %%CITATION = ARXIV:0812.3745;%%
%
%\cite{Campanelli:2009tk}
%\bibitem{Campanelli:2009tk}
  L.~Campanelli,
  %``A Model of Universe Anisotropization,''
  Phys.\ Rev.\  D {\bf 80}, 063006 (2009). 
%  [arXiv:0907.3703 [astro-ph.CO]].
  %%CITATION = PHRVA,D80,063006;%%
%
%%%%%

%%%%%
%\cite{Jimenez:2010hu}
\bibitem{Jimenez:2010hu}
  J.~B.~Jimenez and A.~L.~Maroto,
  %``Cosmological magnetic fields from inflation in extended 
  %electromagnetism,''
  Phys.\ Rev.\  D {\bf 83}, 023514 (2011). 
%  [arXiv:1010.3960 [astro-ph.CO]].
  %%CITATION = PHRVA,D83,023514;%%
%%%%%

\bibitem{NC}
%\cite{Mazumdar:2000jc}
%\bibitem{Mazumdar:2000jc}
  A.~Mazumdar and M.~M.~Sheikh-Jabbari,
  %``Noncommutativity in space and primordial magnetic field,''
  Phys.\ Rev.\ Lett.\  {\bf 87}, 011301 (2001);\ 
%  [arXiv:hep-ph/0012363].
  %%CITATION = PRLTA,87,011301;%%
%\cite{Gamboa:2005bf}
%\bibitem{Gamboa:2005bf}
  J.~Gamboa and J.~Lopez-Sarrion,
  %``U(1) noncommutative gauge fields and magnetogenesis,''
  Phys.\ Rev.\  D {\bf 71}, 067702 (2005). 
%  [arXiv:hep-th/0501034].
  %%CITATION = PHRVA,D71,067702;%%

%\cite{Ashoorioon:2004rs}
\bibitem{Ashoorioon:2004rs}
  A.~Ashoorioon and R.~B.~Mann,
  %``Generation of cosmological seed magnetic fields from inflation with
  %cutoff,''
  Phys.\ Rev.\  D {\bf 71}, 103509 (2005). 
%  [arXiv:gr-qc/0410053].
  %%CITATION = PHRVA,D71,103509;%%

%\cite{Hollenstein:2007kg}
\bibitem{Hollenstein:2007kg}
  L.~Hollenstein, C.~Caprini, R.~Crittenden and R.~Maartens, 
%L.~Hollenstein {\it et al.}, 
  %``Challenges for creating magnetic fields by cosmic defects,''
  Phys.\ Rev.\  D {\bf 77}, 063517 (2008). 
%  [arXiv:0712.1667 [astro-ph]].
  %%CITATION = PHRVA,D77,063517;%%

%%%%%
%\cite{Salim:2006nw}
\bibitem{Salim:2006nw}
  J.~M.~Salim, N.~Souza, S.~E.~Perez Bergliaffa and T.~Prokopec,
  %``Creation of cosmological magnetic fields in a bouncing cosmology,''
  JCAP {\bf 0704}, 011 (2007).
%  [arXiv:astro-ph/0612281].
  %%CITATION = JCAPA,0704,011;%%
%%%%%

%%%%%
%\cite{Maeda:2009hy}
\bibitem{Maeda:2009hy}
  S.~Maeda, S.~Mukohyama and T.~Shiromizu,
  %``Primordial magnetic field from non-inflationary cosmic expansion in
  %Horava-Lifshitz gravity,''
  Phys.\ Rev.\  D {\bf 80}, 123538 (2009). 
%  [arXiv:0909.2149 [astro-ph.CO]].
  %%CITATION = PHRVA,D80,123538;%%
%%%%%

%\cite{Elizalde:2002ca}
\bibitem{Elizalde:2002ca}
  E.~Elizalde, E.~J.~Ferrer and V.~de la Incera,
  %``Beyond constant mass approximation magnetic catalysis in the
gauge Higgs-Yukawa model,''
  Phys.\ Rev.\ D {\bf 68}, 096004 (2003)
  [hep-ph/0209324].
  %%CITATION = HEP-PH/0209324;%%

%\cite{Elizalde:2004mw}
\bibitem{Elizalde:2004mw}
  E.~Elizalde, E.~J.~Ferrer and V.~de la Incera,
  %``Neutrino propagation in a strongly magnetized medium,''
  Phys.\ Rev.\ D {\bf 70}, 043012 (2004)
  [hep-ph/0404234].
  %%CITATION = HEP-PH/0404234;%%

%\cite{Elizalde:2000vz}
\bibitem{Elizalde:2000vz}
  E.~Elizalde, E.~J.~Ferrer and V.~de la Incera,
  %``Neutrino selfenergy and index of refraction in strong magnetic
field: A New approach,''
  Annals Phys.\  {\bf 295}, 33 (2002)
  [hep-ph/0007033].
  %%CITATION = HEP-PH/0007033;%%

%\cite{Elizalde:2012kz}
\bibitem{Elizalde:2012kz}
  E.~Elizalde and V.~Skalozub,
  %``Spontaneous magnetization of the vacuum and the strength of the
magnetic field in the hot Universe,''
  arXiv:1202.3895 [hep-ph], 
to be published in Eur.\ Phys.\ J.\  C.
  %%CITATION = ARXIV:1202.3895;%%

%%%%%
%\cite{Geng:2007va}
\bibitem{Geng:2007va}
  C.~Q.~Geng, S.~H.~Ho and J.~N.~Ng,
  %``Neutrino number asymmetry and cosmological birefringence,''
  JCAP {\bf 0709}, 010 (2007). 
  %[arXiv:0706.0080 [astro-ph]].
  %%CITATION = JCAPA,0709,010;%%
  
\bibitem{GHN2}
% \bibitem{GHN-2}
  C.~Q.~Geng, S.~H.~Ho and J.~N.~Ng,
  %``CPT conserving cosmological birefringence,''
  Int.\ J.\ Mod.\ Phys.\  A {\bf 23}, 3408 (2008); 
%  [arXiv:0801.0024 [astro-ph]].
 Can.\ J.\ Phys.\  {\bf 86}, 587 (2008).
%  [arXiv:0711.4617 [astro-ph]].
  %%CITATION = CJPHA,86,587;%%

\bibitem{BGH}
%\cite{Bamba:2007hf}
%\bibitem{Bamba:2007hf}
  K.~Bamba, C.~Q.~Geng and S.~H.~Ho,
  %``Hypermagnetic Baryogenesis,''
  Phys.\ Lett.\  B {\bf 664}, 154 (2008). 
%  [arXiv:0712.1523 [hep-ph]].
  %%CITATION = PHLTA,B664,154;%%

%%%%%
%\cite{Bamba:2008hr}
\bibitem{Bamba:2008hr}
  K.~Bamba, C.~Q.~Geng and S.~H.~Ho,
  %``Large-scale magnetic fields from inflation due to Chern-Simons-like
  %effective interaction,''
  JCAP {\bf 0811}, 013 (2008). 
%  [arXiv:0806.1856 [astro-ph]].
  %%CITATION = JCAPA,0811,013;%%
%%%%%
%%%%%

%%%%%
%\cite{Campanelli:2008tt}
\bibitem{Campanelli:2008tt}
  L.~Campanelli, P.~Cea and G.~L.~Fogli,
  %``Lorentz Symmetry Violation and Galactic Magnetism,''
  Phys.\ Lett.\  B {\bf 680}, 125 (2009). 
%  [arXiv:0805.1851 [astro-ph]].
  %%CITATION = PHLTA,B680,125;%%

%\cite{Carroll:1989vb}
\bibitem{Carroll:1989vb}
  S.~M.~Carroll, G.~B.~Field and R.~Jackiw,
  %``Limits on a Lorentz and Parity Violating Modification of 
  %Electrodynamics,''
  Phys.\ Rev.\  D {\bf 41}, 1231 (1990).
  %%CITATION = PHRVA,D41,1231;%%
%%%%%

%\cite{Ho:2010aq}
\bibitem{Ho:2010aq}
  S.~H.~Ho, W.~F.~Kao, K.~Bamba and C.~Q.~Geng,
  %``Cosmological birefringence due to CPT-even Chern-Simons-like term with
  %Kalb-Ramond and scalar fields,''
  arXiv:1008.0486 [hep-ph].
  %%CITATION = ARXIV:1008.0486;%%

%\cite{Balaji:2003sw}
\bibitem{Balaji:2003sw}
  K.~R.~S.~Balaji, R.~H.~Brandenberger and D.~A.~Easson,
  %``Spectral dependence of CMB polarization and parity,''
  JCAP {\bf 0312}, 008 (2003). 
%  [arXiv:hep-ph/0310368].
  %%CITATION = JCAPA,0312,008;%%

%%%%%

\bibitem{CMB-anisotropies-Giovannini}
%
%\cite{Giovannini:2009zq}
%\bibitem{Giovannini:2009zq}
  M.~Giovannini,
  %``Estimating relic magnetic fields from CMB temperature correlations,''
  Phys.\ Rev.\  D {\bf 79}, 121302 (2009);\ 
%  [arXiv:0902.4353 [astro-ph.CO]].
  %%CITATION = PHRVA,D79,121302;%%
%
%\cite{Giovannini:2009ts}
%\bibitem{Giovannini:2009ts}
%  M.~Giovannini,
  %``Parameter dependence of magnetized CMB observables,''
%  Phys.\ Rev.\  D {\bf 79}, 103007 (2009)
%\textit{ibid}. 
{\bf 79}, 103007 (2009);\ 
%  [arXiv:0903.5164 [astro-ph.CO]].
  %%CITATION = PHRVA,D79,103007;%%
%
%\cite{Giovannini:2009fu}
%\bibitem{Giovannini:2009fu}
%  M.~Giovannini,
  %``Dark energy, integrated Sachs-Wolfe effect and large-scale magnetic
  %fields,''
  Class.\ Quant.\ Grav.\  {\bf 27}, 105011 (2010). 
%  [arXiv:0907.3235 [astro-ph.CO]].
  %%CITATION = CQGRD,27,105011;%%
%

%%%%%%%%

\bibitem{Kalb-Ramond-field}
%
%\cite{Kar:2000ct}
%\bibitem{Kar:2000ct}
  S.~Kar, P.~Majumdar, S.~SenGupta and A.~Sinha,
  %``Does a Kalb-Ramond field make space-time optically 
  %active?,
  %''
  Eur.\ Phys.\ J.\  C {\bf 23}, 357 (2002);\ 
%  [arXiv:gr-qc/0006097].
  %%CITATION = EPHJA,C23,357;%%
%
%\cite{Kar:2001eb}
%\bibitem{Kar:2001eb}
  S.~Kar, P.~Majumdar, S.~SenGupta and S.~Sur,
  %``Cosmic optical activity from an inhomogeneous 
  %Kalb-Ramond 
  %field,''
  Class.\ Quant.\ Grav.\  {\bf 19}, 677 (2002);\ 
%  [arXiv:hep-th/0109135].
  %%CITATION = CQGRD,19,677;%%
%
%\cite{SenGupta:2002zk}
%\bibitem{SenGupta:2002zk}
  S.~SenGupta and S.~Sur,
  %``Does curvature dilaton coupling with a Kalb-Ramond 
  %field 
  %lead to an
  %accelerating universe?,''
  JCAP {\bf 0312}, 001 (2003);\ 
%  [arXiv:hep-th/0207065].
  %%CITATION = JCAPA,0312,001;%%
%
%\cite{Maity:2003im}
%\bibitem{Maity:2003im}
  D.~Maity and S.~SenGupta,
  %``Cosmic optical activity in a Randall-Sundrum 
  %braneworld 
  %with torsion,''
  Class.\ Quant.\ Grav.\  {\bf 21}, 3379 (2004);\ 
%  [arXiv:hep-th/0311142].
  %%CITATION = CQGRD,21,3379;%%
%
%\cite{Maity:2004yk}
%\bibitem{Maity:2004yk}
  D.~Maity, S.~SenGupta and S.~Sur,
  %``Spinning test particle in Kalb-Ramond background,''
  Eur.\ Phys.\ J.\  C {\bf 42}, 453 (2005);\ 
%  [arXiv:hep-th/0409143].
  %%CITATION = EPHJA,C42,453;%%
%
%\cite{Maity:2004he}
%\bibitem{Maity:2004he}
  D.~Maity, P.~Majumdar and S.~SenGupta,
  %``Parity-violating Kalb-Ramond-Maxwell interactions and 
  %CMB 
  %anisotropy in  a
  %braneworld,''
  JCAP {\bf 0406}, 005 (2004);\ 
%  [arXiv:hep-th/0401218].
  %%CITATION = JCAPA,0406,005;%%
%
%\cite{Maity:2005ah}
%\bibitem{Maity:2005ah}
  D.~Maity, S.~SenGupta and S.~Sur,
  %``Observable signals in a string inspired 
  %axion-dilaton 
  %background and
  %Randall-Sundrum scenario,''
  Phys.\ Rev.\  D {\bf 72}, 066012 (2005). 
%  [arXiv:hep-th/0507210].
  %%CITATION = PHRVA,D72,066012;%%
%
%%%%%

%\bibitem{Observational-Side}
%
%\cite{Ni:2009fg}
\bibitem{Ni:2009fg}
  W.~T.~Ni,
  %``Searches for the role of spin and polarization in gravity,''
  Rept.\ Prog.\ Phys.\  {\bf 73}, 056901 (2010). 
%  [arXiv:0912.5057 [gr-qc]].
  %%CITATION = RPPHA,73,056901;%%
%

%%%
\bibitem{Limits-on-Cosmological-Birefringence}
%
%\cite{Cimatti:1993yc}
%\bibitem{Cimatti:1993yc}
  A.~Cimatti, S.~di Serego Alighieri, G.~B.~Field and R.~A.~E.~Fosbury,
  %``Stellar and scattered light in a radio galaxy at z = 2.63,''
  Astrophys.\ J.\  {\bf 422}, 562 (1994);\ 
  %%CITATION = ASJOA,422,562;%%
%
%\cite{Alighieri:2010eu}
%\bibitem{Alighieri:2010eu}
  S.~d.~S.~Alighieri, F.~Finelli and M.~Galaverni,
  %``Limits on Cosmological Birefringence from the UV Polarization of Distant
  %Radio Galaxies,''
%  Astrophys.\ J.\  {\bf 715}, 33 (2010).
\textit{ibid}. {\bf 715}, 33 (2010). 
%  [arXiv:1003.4823 [astro-ph.CO]].
  %%CITATION = ASJOA,715,33;%%
%
%%%

\bibitem{C-H}
%\cite{Rubakov:1982df}
%\bibitem{Rubakov:1982df}
  V.~A.~Rubakov, M.~V.~Sazhin and A.~V.~Veryaskin,
  %``Graviton Creation In The Inflationary Universe And The Grand Unification
  %Scale,''
  Phys.\ Lett.\  B {\bf 115}, 189 (1982);\
  %%CITATION = PHLTA,B115,189;%%
%\cite{Abbott:1984fp}
%\bibitem{Abbott:1984fp}
  L.~F.~Abbott and M.~B.~Wise,
  %``Constraints On Generalized Inflationary Cosmologies,''
  Nucl.\ Phys.\  B {\bf 244}, 541 (1984).
  %%CITATION = NUPHA,B244,541;%%

%\cite{Komatsu:2010fb}
\bibitem{Komatsu:2010fb}
  E.~Komatsu {\it et al.}  [WMAP Collaboration],
  %``Seven-Year Wilkinson Microwave Anisotropy Probe (WMAP) Observations:
  %Cosmological Interpretation,''
  Astrophys.\ J.\ Suppl.\  {\bf 192}, 18 (2011). 
%  [arXiv:1001.4538 [astro-ph.CO]].
  %%CITATION = APJSA,192,18;%%

\bibitem{Kolb and Turner}
E.~W.~Kolb and M.~S.~Turner,
\textit{The Early Universe}
(Addison-Wesley, Redwood City, California, 1990).

%%%%%
\bibitem{YIKM}
%
%\cite{Yamazaki:2008jh}
%\bibitem{Yamazaki:2008jh}
  D.~G.~Yamazaki, K.~Ichiki, T.~Kajino and G.~J.~Mathews,
  %``Constraints on the Primordial Magnetic Field from $\sigma_8$,''
  Phys.\ Rev.\  D {\bf 78}, 123001 (2008);\ 
%  [arXiv:0811.2221 [astro-ph]].
  %%CITATION = PHRVA,D78,123001;%%
%
%\cite{Yamazaki:2010jw}
%\bibitem{Yamazaki:2010jw}
%  D.~G.~Yamazaki, K.~Ichiki, T.~Kajino and G.~J.~Mathews,
  %``Constraints on the neutrino mass and the primordial magnetic field 
  %from the
  %matter density fluctuation parameter $\sigma_8$,''
%  Phys.\ Rev.\  D {\bf 81}, 103519 (2010)
\textit{ibid}. {\bf 81}, 103519 (2010). 
%  [arXiv:1005.1638 [astro-ph.CO]].
  %%CITATION = PHRVA,D81,103519;%%
%
%%%%%

%%%%%
%\cite{Wang:2008vp}
\bibitem{Wang:2008vp}
  S.~Wang,
  %``New primordial-magnetic-field limit from the latest LIGO S-5 data,''
  Phys.\ Rev.\  D {\bf 81}, 023002 (2010). 
%  [arXiv:0810.5620 [astro-ph]].
  %%CITATION = PHRVA,D81,023002;%%
%%%%%

%%%%%
%\cite{Tashiro:2010st}
\bibitem{Tashiro:2010st}
  H.~Tashiro, K.~Takahashi and K.~Ichiki,
  %``Primordial magnetic fields with X-ray and S-Z cluster survey,''
  arXiv:1010.4407 [astro-ph.CO].
  %%CITATION = ARXIV:1010.4407;%%
%%%%%

%%%%%
%\cite{Bamba:2007hm}
\bibitem{Bamba:2007hm}
  K.~Bamba,
  %``Property of the spectrum of large-scale magnetic fields from inflation,''
  Phys.\ Rev.\  D {\bf 75}, 083516 (2007). 
%  [arXiv:astro-ph/0703647].
  %%CITATION = PHRVA,D75,083516;%%
%%%%%

%\cite{Freedman:2000cf}
\bibitem{Freedman:2000cf}
  W.~L.~Freedman {\it et al.}  [HST Collaboration],
  %``Final Results from the Hubble Space Telescope Key Project to Measure the
  %Hubble Constant,''
  Astrophys.\ J.\  {\bf 553}, 47 (2001).
%  [arXiv:astro-ph/0012376].
  %%CITATION = ASJOA,553,47;%%

\bibitem{BBN}
%\cite{Grasso:1996kk}
%\bibitem{Grasso:1996kk}
  D.~Grasso and H.~R.~Rubinstein,
  %``Revisiting Nucleosynthesis Constraints on Primordial Magnetic Fields,''
  Phys.\ Lett.\  B {\bf 379}, 73 (1996);\
%  [arXiv:astro-ph/9602055].
  %%CITATION = PHLTA,B379,73;%%
%\cite{Cheng:1996yi}
%\bibitem{Cheng:1996yi}
  B.~l.~Cheng, A.~V.~Olinto, D.~N.~Schramm and J.~W.~Truran,
  %``Constraints on the strength of primordial magnetic fields from big bang
  %nucleosynthesis revisited,''
  Phys.\ Rev.\  D {\bf 54}, 4714 (1996).
%  [arXiv:astro-ph/9606163].
  %%CITATION = PHRVA,D54,4714;%%

\bibitem{CMB-Limit}
%\bibitem{Subramanian}
%\cite{Subramanian:1998fn}
%\bibitem{Subramanian:1998fn}
  K.~Subramanian and J.~D.~Barrow,
  %``Microwave Background Signals from Tangled Magnetic Fields,''
  Phys.\ Rev.\ Lett.\  {\bf 81}, 3575 (1998);\
%  [arXiv:astro-ph/9803261].
  %%CITATION = PRLTA,81,3575;%%
%\cite{Seshadri:2000ky}
%\bibitem{Seshadri:2000ky}
  T.~R.~Seshadri and K.~Subramanian,
  %``CMBR Polarization Signals from Tangled Magnetic Fields,''
%  Phys.\ Rev.\ Lett.\  {\bf 87}, 101301 (2001)
\textit{ibid}. {\bf 87}, 101301 (2001);\
%  [arXiv:astro-ph/0012056].
  %%CITATION = PRLTA,87,101301;%%
%\cite{Subramanian:2002nh}
%\bibitem{Subramanian:2002nh}
  K.~Subramanian and J.~D.~Barrow,
  %``Small-scale microwave background anisotropies due to tangled primordial
  %magnetic fields,''
  Mon.\ Not.\ Roy.\ Astron.\ Soc.\  {\bf 335}, L57 (2002);\
%  [arXiv:astro-ph/0205312].
  %%CITATION = MNRAA,335,L57;%%
%\cite{Subramanian:2003sh}
%\bibitem{Subramanian:2003sh}
  K.~Subramanian, T.~R.~Seshadri and J.~D.~Barrow,
  %``Small-scale CMB polarization anisotropies due to tangled primordial
  %magnetic fields,''
%  Mon.\ Not.\ Roy.\ Astron.\ Soc.\  {\bf 344}, L31 (2003)
\textit{ibid}. {\bf 344}, L31 (2003);\ 
%  [arXiv:astro-ph/0303014].
  %%CITATION = MNRAA,344,L31;%%
%
%\cite{Tashiro:2005hc}
%\bibitem{Tashiro:2005hc}
  H.~Tashiro, N.~Sugiyama and R.~Banerjee,
  %``Nonlinear Evolution of Cosmic Magnetic Fields and Cosmic Microwave
  %Background Anisotropies,''
  Phys.\ Rev.\  D {\bf 73}, 023002 (2006);\ 
%  [arXiv:astro-ph/0509220].
  %%CITATION = PHRVA,D73,023002;%%
%
%\cite{Yamazaki:2008gr}
%\bibitem{Yamazaki:2008gr}
  D.~G.~Yamazaki, K.~Ichiki, T.~Kajino and G.~J.~Mathews,
  %``Effects of a Primordial Magnetic Field on Low and High Multipoles of the
  %CMB,''
%  Phys.\ Rev.\  D {\bf 77}, 043005 (2008);\ 
\textit{ibid}. {\bf 77}, 043005 (2008);\ 
%  [arXiv:0801.2572 [astro-ph]].
  %%CITATION = PHRVA,D77,043005;%%
%
%\cite{Giovannini:2008df}
%\bibitem{Giovannini:2008df}
  M.~Giovannini and K.~E.~Kunze,
  %``CMB polarization induced by stochastic magnetic fields,''
  arXiv:0804.2238 [astro-ph];\ 
  %%CITATION = ARXIV:0804.2238;%%
%
%\cite{Kahniashvili:2008hx}
%\bibitem{Kahniashvili:2008hx}
  T.~Kahniashvili, Y.~Maravin and A.~Kosowsky,
  %``Faraday Rotation Limits On A Primordial Magnetic Field From Wilkinson
  %Microwave Anisotropy Probe Five-Year Data,''
  Phys.\ Rev.\  D {\bf 80}, 023009 (2009);\ 
%  [arXiv:0806.1876 [astro-ph]].
  %%CITATION = PHRVA,D80,023009;%%
%
%\cite{Yamazaki:2010nf}
%\bibitem{Yamazaki:2010nf}
  D.~G.~Yamazaki, K.~Ichiki, T.~Kajino and G.~J.~Mathews,
  %``New Constraints on the Primordial Magnetic Field,''
%  Phys.\ Rev.\  D {\bf 81}, 023008 (2010)
\textit{ibid}. {\bf 81}, 023008 (2010);\ 
%  [arXiv:1001.2012 [astro-ph.CO]].
  %%CITATION = PHRVA,D81,023008;%%
%
%\cite{Shaw:2010ea}
%\bibitem{Shaw:2010ea}
  J.~R.~Shaw and A.~Lewis,
  %``Constraining Primordial Magnetism,''
  arXiv:1006.4242 [astro-ph.CO];\ 
  %%CITATION = ARXIV:1006.4242;%%
%
%\cite{Trivedi:2010gi}
%\bibitem{Trivedi:2010gi}
  P.~Trivedi, K.~Subramanian and T.~R.~Seshadri,
  %``Primordial Magnetic Field Limits from Cosmic Microwave Background
  %Bispectrum of Magnetic Passive Scalar Modes,''
  Phys.\ Rev.\  D {\bf 82}, 123006 (2010);\ 
%  [arXiv:1009.2724 [astro-ph.CO]].
  %%CITATION = PHRVA,D82,123006;%%
%
%\cite{Shiraishi:2010yk}
%\bibitem{Shiraishi:2010yk}
  M.~Shiraishi, D.~Nitta, S.~Yokoyama, K.~Ichiki and K.~Takahashi,
  %``Cosmic microwave background bispectrum of vector modes induced from
  %primordial magnetic fields,''
  Phys.\ Rev.\  D {\bf 82}, 121302 (2010)
  [Erratum-ibid.\  D {\bf 83}, 029901 (2011)]
  [Phys.\ Rev.\  D {\bf 83}, 029901 (2011)]. 
%  [arXiv:1009.3632 [astro-ph.CO]].
  %%CITATION = PHRVA,D83,029901;%%
%

%\cite{Barrow:1997mj}
\bibitem{Barrow:1997mj}
  J.~D.~Barrow, P.~G.~Ferreira and J.~Silk,
  %``Constraints on a Primordial Magnetic Field,''
  Phys.\ Rev.\ Lett.\  {\bf 78}, 3610 (1997).
%  [arXiv:astro-ph/9701063].
  %%CITATION = PRLTA,78,3610;%%

%%%%% Detect %%%%%

\bibitem{Planck-1}
 See http://www.sciops.esa.int/index.php?project=PLANCK.

\bibitem{Planck-2}
http://www.rssd.esa.int/SA/PLANCK/docs/Bluebook-ESA-SCI(2005)1\_V2.pdf.

\bibitem{QUIET-1}
 See http://quiet.uchicago.edu/index.php.

%\cite{Samtleben:2008rb}
\bibitem{Samtleben:2008rb}
  D.~Samtleben and f.~t.~Q.~Collaboration,
  %``Measuring the Cosmic Microwave Background Radiation (CMBR) polarization
  %with QUIET,''
  Nuovo Cim.\  {\bf 122B}, 1353 (2007)
  [arXiv:0802.2657 [astro-ph]].
  %%CITATION = NUCIA,122B,1353;%%

\bibitem{B-Pol}
 See http://www.b-pol.org/index.php.

\bibitem{LiteBIRD}
 See http://cmbpol.kek.jp/litebird/.

\bibitem{Test}
%
%\cite{Caprini:2003vc}
%\bibitem{Caprini:2003vc}
  C.~Caprini, R.~Durrer and T.~Kahniashvili,
  %``The Cosmic Microwave Background and Helical Magnetic Fields: the tensor
  %mode,''
  Phys.\ Rev.\  D {\bf 69}, 063006 (2004);\ 
%  [arXiv:astro-ph/0304556].
  %%CITATION = PHRVA,D69,063006;%%
%
%\cite{Kahniashvili:2005xe}
%\bibitem{Kahniashvili:2005xe}
  T.~Kahniashvili and B.~Ratra,
  %``Effects of Cosmological Magnetic Helicity on the Cosmic Microwave
  %Background,''
%  Phys.\ Rev.\  D {\bf 71}, 103006 (2005).
\textit{ibid}. {\bf 71}, 103006 (2005);\ 
%  [arXiv:astro-ph/0503709].
  %%CITATION = PHRVA,D71,103006;%%
%
%\cite{Kahniashvili:2006zs}
%\bibitem{Kahniashvili:2006zs}
  T.~Kahniashvili,
  %``Effects of primordial helicity on CMB,''
  New Astron.\ Rev.\  {\bf 50}, 1015 (2006);\ 
%  [arXiv:astro-ph/0605440].
  %%CITATION = ASTRE,50,1015;%%
%
%\cite{Kristiansen:2008tx}
%\bibitem{Kristiansen:2008tx}
  J.~R.~Kristiansen and P.~G.~Ferreira,
  %``Constraining primordial magnetic fields with CMB polarization
  %experiments,''
  Phys.\ Rev.\  D {\bf 77}, 123004 (2008).
%  [arXiv:0803.3210 [astro-ph]].
  %%CITATION = PHRVA,D77,123004;%%
%

%%%%%%%%%%%%%%%%%%

\end{thebibliography}
\end{document}